\documentclass[aps, reprint, amsmath,amssymb, prx, superscriptaddress]{revtex4-2}

\usepackage{svg}
\usepackage{color}
\usepackage{graphicx}
\usepackage{dcolumn}
\usepackage{bm}
\usepackage{physics}
\usepackage{hyperref}
\usepackage{url}
\usepackage{amsmath}
\usepackage{notes2bib}
\usepackage{ dsfont }
\usepackage{svg}
\usepackage{verbatim}
\usepackage{amssymb}
\usepackage{amsfonts}              
\usepackage{amsthm}                
\usepackage{mathtools}
\usepackage{kotex}
\usepackage{proof-at-the-end}
\usepackage{qcircuit}
\usepackage[normalem]{ulem}
\usepackage{xcolor}

\DeclarePairedDelimiterX{\barpair}[2]{(}{)}{%
  #1\;\delimsize\|\;#2%
}

\newtheorem{theorem}{Theorem}

\newtheorem{lemma}[theorem]{Lemma}

\newtheorem{corollary}[theorem]{Corollary}

\NewDocumentEnvironment{thmE}{O{}O{}+b}{%
    \begin{theoremEnd}[normal,#2]{thm}[#1]%
        #3%
    \end{theoremEnd}%
    }{}
    \NewDocumentEnvironment{proofE}{O{}+b}{%
        \begin{proofEnd}[#1]%
             #2%
        \end{proofEnd}%
}{}

\bibnotesetup{note-name    = ,use-sort-key = false} 
\begin{document}

\title{Correlational Resource Theory of\\Catalytic Quantum Randomness under Conservation Law}

\author{Seok Hyung Lie}
\author{Hyunseok Jeong}
\affiliation{%
 Department of Physics and Astronomy, Seoul National University, Seoul, 151-742, Korea
}%

\date{\today}

\begin{abstract}
Catalysts are substances that assist transformation of other resourceful objects without being consumed in the process. However, the fact that their `catalytic power' is limited and can be depleted is often overlooked, especially in the recently developing theories on catalysis of quantum randomness utilizing building correlation with catalyst.  In this work, we establish a resource theory of one-shot catalytic randomness in which uncorrelatedness is consumed in catalysis of randomness. We do so by completely characterizing bipartite unitary operators that can be used to implement catalysis of randomness using partial transpose. By doing so, we find that every catalytic channel is factorizable, and therefore there exists a unital channel that is not catalytic. We define a family of catalytic entropies that quantifies catalytically extractable entropy within a quantum state and show how much degeneracy of quantum state can boost the catalytic entropy beyond its ordinary entropy. Based on this, we demonstrate that a randomness source can be actually exhausted after a certain amount of randomness is extracted. We apply this theory to systems under conservation law that forbids superposition of certain quantum states and find that non-maximally mixed states can yield the maximal catalytic entropy. We discuss implications of this theory to various topics including catalytic randomness absorption, the no-secret theorem and the possibility of multi-party infinite catalysis.
\end{abstract}

\pacs{Valid PACS appear here}
\maketitle

\section{Introduction}
A catalyst is a substance that accelerates or initiates chemical reactions without being consumed or destroyed. This concept has been adopted in the context of quantum information for manipulation of entanglement, coherence and realization of thermal operations. Recently, a generalized concept of catalytic randomness for state transitions has been explored \cite{boes2018catalytic, boes2019neumann, lie2019unconditionally, lie2020minent, lie2020randomness, muller2018correlating}. In this generalized setting, a randomness source, a mixed quantum state that serves as a source of randomness for otherwise deterministic process, is catalytically used in the sense that its state remains unchanged after the interaction taking place. However, the randomness source, as a catalyst, is allowed to be correlated with other quantum systems in the course of interaction so that immediate recycle of the catalyst is impossible for the interaction with the very same quantum system it interacted with. Therefore, this concept of `catalysis' has a certain limitation.

This situation has the following everyday analogy. Consider a transaction between two parties and coins used as currency for it. The coins per se, as a physical manifestation of economic value, are not deteriorated or consumed in the transactional process, but their relation with other agents is changed, i.e. their ownership is transferred. This is the very reason why one person cannot buy an indefinite amount of products even when no actual coin or bill is consumed or destroyed; one has only a limited amount of money owned by herself.

A more direct example in information theory of such phenomena is one-time pad. One-time pad is a table of random numbers that can be used for secure cryptographic communication. Note that the table itself remains  intact and random for someone who never interacted with it, but a user cannot use the same table twice lest the communication becomes insecure, hence the name `one-time pad'. These observations motivate the explicit identification and treatment of this relational resource being consumed in information processing processes.

In this work, we set to establish such a theory for catalytic randomness for implementing quantum channels. We identify uncorrelatedness is the resource being consumed in catalysis, and show that randomness produced in the process is extracted from such uncorrelatedness. As a result, we define a quantity called \textit{catalytic entropy} for arbitrary quantum state, which equals to the maximal amount of entropy that can be extracted from the quantum state through catalysis. A significant consequence is that a randomness source correlated enough with the user can be \textit{depleted}. This perspective on randomness aligns with more conventional resource theories in quantum information science in which a resource has extensive quantity that can be produced or consumed.

We also generalize the theory of catalytic quantum randomness significantly. First, we characterize the bipartite unitary operators that are still unitary after partial transposition as catalysis unitary operators. For this purpose, we show that every catalysis unitary is compatible with maximally mixed catalyst, and show that catalysis unitaries should have the controlled-unitary form with they are compatible with non-uniform catalysts. We also discuss about catalysts given in an already correlated form and randomness deposit through catalysis. Next, we introduce a few advantages of the approach that treats the correlation formed between the system and catalyst explicitly, including the infinite multi-party catalysis. In doing so, we show that the partial transpose of a catalysis unitary has an operational meaning as the recovery map that recovers the input state of the catalysis encoded in the correlation with environment.

\section{Notations} \label{sec:notations}
We will denote the marginal state of a multipartite quantum state $\rho_{ABC\dots}$ on the system $A$ as $\rho_A$. However, the system subscripts will be omitted when it is obvious from context. We will frequently use the von Neumann entropy of quantum state $\rho$ defined as $S(\rho):=-\Tr[\rho\log_2 \rho]$. When the notation such as $S(A)_\rho$ is used, it represents the von Neumann entropy of the marginal state $\rho_A$ of the multipartite state $\rho_{ABC\dots},$ i.e. $S(A)_\rho=S(\rho_A)$. These two notations will be used interchangeably. Similarly Shannon entropy $H(p):=-\sum_i p_i \log_2  p_i$ and R\'{e}nyi entropy \cite{renyi1961measures}, $H_\alpha(p)=\frac{1}{1-\alpha}\log_2 \sum_i p_i^\alpha$ are defined for probability distributions $\{p_i\}$. A quantum channel, or a quantum map, is a linear map on a operator space that that is completely positive and trace preserving. A unital quantum map is a quantum map that preserves the maximally mixed state. We will denote the dimension of the Hilbert space associated with system $S$ as $d_S$ from now on, with the exception that the dimension of the Hilbert space associated with the input state $\rho$ being denoted as just $d$. A $d_A \otimes d_B$-dimensional Hilbert space stands for the tensor product of $d_A$-dimensional and $d_B$-dimensional Hilbert spaces.

\section{main result}

\subsection{Catalytic Randomness}\label{subsec:catran}

 Assume that there exists a deterministic agent $A$ who cannot generated randomness without interacting with an outer agent. In quantum mechanics, it means the only actions $A$ can take aside from appending auxiliary systems are unitary operations. Suppose that $A$ is allowed to borrow a system $B$ called \textit{catalyst} in the quantum state $\sigma_B$ to implement a quantum map $\Phi$. $A$ is allowed to interact with $B$ but should return the system $B$ in its original state $\sigma_B$ after every interaction. This can be summarized as the following two conditions. When a bipartite unitary $U$ on systems $A$ and $B$ is used to implement a quantum map $\rho \mapsto \Phi(\rho)$ with a catalyst $\sigma$, i.e.
 
 \begin{figure}[t]
    \includegraphics[width=.5\textwidth]{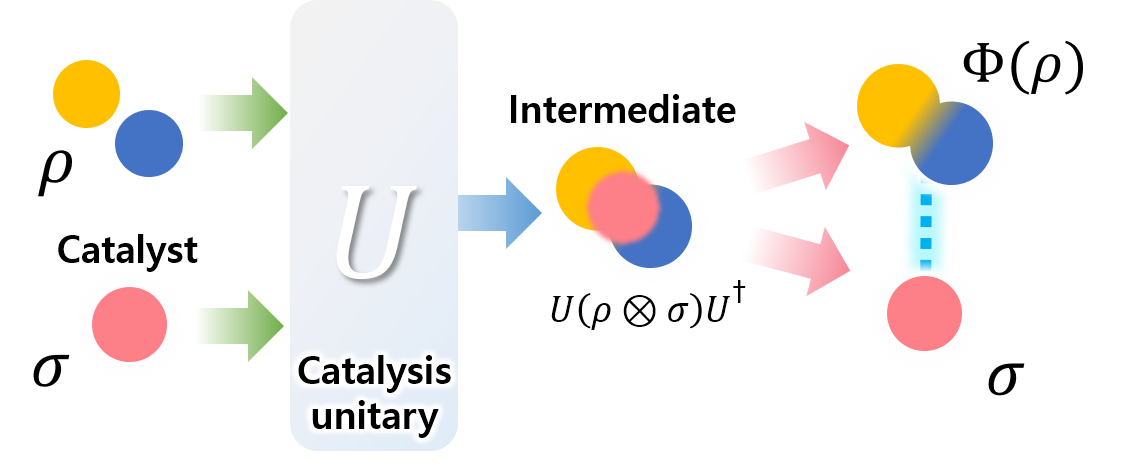}
    \caption{Schematic depiction of catalysis process. Catalyst $\sigma$ is used to implement the quantum map $\rho \mapsto \Phi(\rho)$.  The catalyst stays in its original form $\sigma$ as the marginal state of the global state $U(\rho\otimes\sigma)U^\dag$ called the intermediate, after the interaction, regardless of the input state $\rho$. The blue dotted line depicts the correlation formed between the system and the catalyst, which indicates that the free randomness in the catalyst is used during the process.}\label{fig:catalysis}
\end{figure}

\begin{equation}
    \Tr_B U(\rho_{A} \otimes \sigma_{B})U^\dag = \Phi(\rho).
\end{equation}
The catalyst $\sigma$ should retain its original randomness, i.e. spectrum, after the interaction regardless of the input state $\rho$, i.e.
\begin{equation} \label{eqn:catal}
    \Tr_A U(\rho_{A} \otimes \sigma_{B})U^\dag = V\sigma_{B}V^\dag, \quad\forall \rho,
\end{equation}
with some unitary operator $V$ on the system $B$. We will call a quantum map implemented in this manner or the bipartite interaction as a whole a \textit{catalysis} or a \textit{catalysis process} depending on the context, and call the bipartite unitary operator used for catalysis a \textit{catalysis unitary} operator.

We will say that $U$ is compatible with $\sigma$ if (\ref{eqn:catal}) holds and vice versa.  For the sake of convenience, we will often use the definition of the compatibility for the cases where $\sigma_B$ is an unnormalized Hermitian operator, too. Similar randomness-utilizing processes were considered in previous works, under the name noisy operations \cite{horodecki2003reversible, scharlau2018quantum, gour2015resource} or thermal operations. However, most studies were focused on the implementation of the transition between two fixed quantum states and the existence of a feasible catalyst for that task. Here, we are more interested in the implementation of quantum map, independently of potential input state, with a given catalyst.

The condition (\ref{eqn:catal}) may look too strong, but actually it is equivalent to an apparently weaker conditions. We refine the result of Ref. \cite{lie2020uniform} to get the following equivalent conditions. All the omitted proofs of results can be found in Appendix \ref{appendix}.
\begin{theoremEnd}{prop} \label{prop:charac}
    For any bipartite unitary operator $U$ on a composite system $AB$, the following requirements are equivalent:
    \begin{equation*}
       (i)\quad \Tr_A U(\rho_{A} \otimes \sigma_{B})U^\dag = V\sigma_{B}V^\dag, \quad\forall \rho,
    \end{equation*}
    with some unitary operator $V$ on $B$. 
    \begin{equation*}
        (ii) \quad \Tr_A U(\rho_{A} \otimes \sigma_{B})U^\dag = W_\rho\sigma_{B}W_\rho^\dag, \quad\forall \rho,
    \end{equation*}
    with some unitary operator $W_\rho$ on $B$ depending on $\rho$. 
    \begin{equation*}
        (iii) \quad \Tr_A U(\rho_{A} \otimes \sigma_{B})U^\dag = \xi_B, \quad\forall \rho,
    \end{equation*}
    for some quantum state $\xi_B$ on $B$ independent of $\rho_A$.
\end{theoremEnd}
\begin{proofEnd}
    $(i)\Rightarrow(ii)$ is trivial. $(ii)\Rightarrow(i)$ can be proved as follows. Consider a convex combination of two arbitrary inputs $\rho_1$ and $\rho_2,$ i.e. $\tau=\frac{1}{2}(\rho_1+\rho_2)$. Then, from the invariance of the von Neumann entropy under unitary transformation, $S(W_{\tau}\sigma W_{\tau}^\dag))=S(\sigma)=\frac{1}{2}(S(W_{\rho_1}\sigma W_{\rho_1}^\dag) +S(W_{\rho_2}\sigma W_{\rho_2}^\dag))$. Note that, from the linearity, it follows that $W_\tau\sigma W_\tau^\dag=\frac{1}{2}(W_{\rho_1}\sigma W_{\rho_1}^\dag +W_{\rho_2}\sigma W_{\rho_2}^\dag)$. Therefore, from the saturation condition of the subadditivity of the von Neumann entropy \cite{nielsen2002quantum}, it follows that $W_{\rho_1}\sigma W_{\rho_1}^\dag=W_{\rho_2}\sigma W_{\rho_2}^\dag$. The equivalence of $(i)$ and $(iii)$ was shown in Ref. \cite{lie2020uniform}.
\end{proofEnd}
The condition $(ii)$ states that, when considering $B$ as a thermal bath, the process is \textit{adiabatic} in the sense that the bath undergoes no change of entropy. The change of entropy of the system $A$ is still allowed and can only be positive, as it is for free expansion of ideal gas. The requirement $(iii)$ gives a characterization that catalytic quantum map is a quantum map that can be implemented without leaking any information of input state to the ancillary system. These observations put catalysis of quantum randomness in the context of various research topics including quantum thermodynamics, private quantum decoupling \cite{buscemi2009private} and quantum secret sharing \cite{gottesman2000theory,cleve1999share,imai2005information}.

Because of Theorem \ref{prop:charac}, for every catalysis with the catalysis unitary $U$, we have corresponding unitary operator $V$ on $B$ in (\ref{eqn:catal}). We can consider a new catalysis unitary $(\mathds{1}_A \otimes V_B^\dag)U$ that completely preserves its catalyst, e.g. $\sigma \to \sigma$ while implementing the same quantum channel. We will call such a form of a catalysis unitary its \textit{canonical form}. 

Before developing a resource theoretical approach to catalytic randomness, we first show that every catalysis unitary is compatible with the maximally mixed states. It means that for arbitrary catalysis, even when one replaces the catalyst with the maximally mixed state, it will still be a catalysis.

\begin{theoremEnd}{prop} \label{prop:comp}
    A catalysis unitary compatible with a catalyst $\sigma$ is also compatible with its normalized projection onto its eigenspaces. Furthermore, every catalysis unitary is compatible with the maximally mixed catalyst.
\end{theoremEnd}
\begin{proofEnd}

 The following lemma was first proved as a special case more general result for von Neumann algebra theory \cite{arias2002fixed, lie2020uniform}. Here we give a more elementary proof.

\begin{lemma} \label{lem:unital}
Let $\Phi$ be a unital channel on a finite dimensional Hilbert space $\mathcal{H}$ given as $\Phi(\rho):=\sum_i K_i\rho K_i^\dag$. If $\Phi$ fixes a positive Hermitian operator $\sigma > 0$ on $\mathcal{H}$ i.e. $\Phi(\sigma)=\sigma$, then $\Phi$ also fixes the projector onto each eigenspace of $\sigma$. Furthermore, each projector commutes with each Kraus operator $K_i$ of $\Phi$ regardless of the choice of Kraus operators.
\end{lemma}
\begin{proof}
    Without loss of generality, we can assume that $\sigma$ has at least two different eigenvalues. Let $\lambda_i$ be the $i$-th largest eigenvalue of $\sigma$ with $\Pi_i$ being the projector onto the corresponding eigenspace. We first prove that $\Phi$ fixes the projector $\Pi_m$ onto the eigenspace corresponding to the smallest eigenvalue $\lambda_m$ of $\sigma$. It will prove the desired lemma since, then, $\Phi$ also fixes $\sigma+\|\sigma\|\Pi_m$ whose smallest eigenvalue is the second smallest eigenvalue of $\sigma$ and the same conclusion can be drawn about $\sigma+\|\sigma\|\Pi_m$.
    First, let $\ket{\psi}$ be an arbitrary eigenvector of $\sigma$ corresponding to $\lambda_m$. We conjugate $\Phi(\sigma)=\sigma$ with $\ket{\psi}$ to get the following equation.
    \begin{equation} \label{eqn:avglam}
        \lambda_m=\sum_i \bra{\psi}\Phi(\Pi_i)\ket{\psi}\lambda_i.
    \end{equation}
    Here, $\{\bra{\psi}\Phi(\Pi_i)\ket{\psi}\}_{i=1}^m$ forms a probability distribution since $\sum_i \Pi_i = \mathds{1}$ and $\Phi$ is unital. Therefore, the right hand side of (\ref{eqn:avglam}) is an average of $\{\lambda_i\}$, which is strictly larger than $\lambda_m$ whenever $\bra{\psi}\Phi(\Pi_m)\ket{\psi}<1$. Therefore we have $\bra{\psi}\Phi(\Pi_m)\ket{\psi}=1$. Since this result holds for arbitrary eigenvector $\ket{\psi}$ corresponding $\lambda_m$, we have $\Phi(\Pi_m)=\Pi_m \oplus P$ for some $P\geq 0$, but since $\Tr\Phi(\Pi_m)=\Tr(\Pi_m)$, we have $\Phi(\Pi_m)=\Pi_m$.
    
    Let $\ket{\psi}$ and $\ket{\phi}$ be eigenvectors corresponding to distinct eigenvalues $\lambda_r$ and $\lambda_s$ of $\sigma$. Then we have
    \begin{equation}
        \bra{\psi}\Phi(\dyad{\phi})\ket{\psi}\leq \bra{\psi}\Pi_s\ket{\psi}=0,
    \end{equation}
    so we have $\bra{\psi}\Phi(\dyad{\phi})\ket{\psi}=0$ but $\bra{\psi}\Phi(\dyad{\phi})\ket{\psi}=\sum_i \bra{\psi}K_i\dyad{\phi}K_i^\dag\ket{\psi}=\sum_i |\bra{\psi}K_i\ket{\phi}|^2$. It implies that $\bra{\psi}K_i\ket{\phi}=0$ for every $i$, which implies that $\left[\Pi_i,K_j\right]=0$ for every $i$ and $j$.
\end{proof}
Lemma \ref{lem:unital} yields the following result. Without loss of generality, we assume that the catalysis unitary $U$ is in its canonical form. Consider the quantum channel $T$ defined as  $T(\tau):=\Tr_A U(\frac{1}{d}\mathds{1}\otimes \tau)U^\dag$. Note that $T$ is a unital channel that also fixes $\sigma$. Therefore, if $\{\ket{s}\}$ is a basis on $A$, then $\frac{1}{\sqrt{d}}(\bra{s}\otimes \mathds{1})U(\ket{r}\otimes\mathds{1})$, Kraus operators of $T$, commute with $\Pi_i$, arbitrary projector onto one of eigenspaces of $\sigma$. Therefore the catalysis unitary operator $U$ itself also commutes with every $\mathds{1}\otimes \Pi_i$. It implies that $\Pi_i$ are also compatible with $U$ since
   \begin{align}
       \lambda_i \Tr_A U(\rho \otimes \Pi_i)U^\dag =& \Tr_A U(\rho \otimes \Pi_i \sigma \Pi_i)U^\dag \\
       =&\Pi_i \Tr_A U(\rho \otimes \sigma)U^\dag \Pi_i \\
       =&\Pi_i \sigma \Pi_i=\lambda_i \Pi_i,
   \end{align}
   for arbitrary $\rho$. By the linearity, it follows that $\sum_i \Pi_i=\mathds{1}_B$ is also compatible with $U$. 
\end{proofEnd}
Proposition \ref{prop:comp} shows that any catalytic map $\Phi$ implemented with a catalysis unitary $U$ on $\mathcal{H}_A\otimes\mathcal{H}_B$ with a catalyst $\sigma_B$ with the spectral decomposition $\sigma=\sum_i\lambda_i\Pi_i$ $(\Pi_i\Pi_j=\delta_{ij}\Pi_i)$ can be decomposed into sub-catalyses. To be more precise, if $\mathcal{H}_i$ is the support of $\Pi_i$, then one can decompose the Hilbert space $\mathcal{H}_B=\bigoplus_i \mathcal{H}_i$ and the catalysis unitary  $U=\bigoplus_i U_i$ where $U_i$ is defined on $\mathcal{H}_A\otimes\mathcal{H}_i$. Let $r_i=\Tr\Pi_i$ and $\pi_i=r_i^{-1}\Pi_i$. Then, we get that $\Phi$ is a convex sum of other catalytic maps that uses a maximally mixed state as its catalyst, i.e., $\Phi=\sum_i \lambda_i r_i \Phi_i$ where $\Phi_i(\rho)= \Tr_{\mathcal{H}_i} U_i(\rho_A\otimes\pi_i) U_i^\dag$.

The unital maps that can be implemented with a finite dimensional quantum system prepared in the maximally mixed state  as its ancillary system are known as the \textit{exactly factorizable maps} \cite{shor2010structure,haagerup2011factorization}, which is in turn a special case of more general \textit{factorizable maps}, whose ancillary systems can be represented with a (possibly infinite dimensional) von Neumann algebra. The catalytic maps $\Phi_i$ defined above are therefore factorizable maps, but, since the set of factorizable maps is known to be convex, we can see that arbitrary catalytic map is also factorizable. However, since there are non-factorizable unital maps \cite{haagerup2011factorization}, we get the following results.

\begin{theoremEnd}{theorem} \label{thm:factorizable}
    Not every unital map is catalytic. 
\end{theoremEnd}
Theorem \ref{thm:factorizable} solves an open problem introduced in Ref.\cite{lie2020uniform}, which asked the exact inclusion relation of the set of unital maps and the set of catalytic maps. In light of Proposition \ref{prop:charac}, it follows that there is a unital quantum map that must leak some information of the input system to \textit{whatever} system coupled with the input system to implement the quantum map. This result is rather surprising, because even the erasure map, which completely deletes the information of input state, can be implemented without leaking any information to an ancillary system.

Using Proposition \ref{prop:comp}, we can also completely characterize the class of catalysis unitary operators.

\begin{theoremEnd}{theorem} \label{thm:catauni}
    A bipartite unitary operator $U$ on two systems $A$ and $B$ is a catalysis unitary operator if and only if its partial transpose $U^{T_A}$ is also a unitary operator.
\end{theoremEnd}
\begin{proofEnd}
    First, assume that $U:\mathcal{H}_{AB}\to\mathcal{H}_{AB}$ a unitary operator whose partial transpose $U^{T_A}$ is also unitary. We define a unnormalized maximally entangled  state on system $A$ and its copy $A'$ as $\ket{\Gamma}:=\sum_i \ket{ii}_{AA'}$. Then, for any quantum state on system $A$,
    \begin{gather}
        \Tr_A U(\rho_A \otimes \frac{1}{d_B}\mathds{1}_B)U^\dag\\
        =\bra{\Gamma}_{AA'} U_{AB}(\rho_A \otimes \frac{1}{d_B}\mathds{1}_B)U_{AB}^\dag \ket{\Gamma}_{AA'}\\
        =\bra{\Gamma}_{AA'}  U_{A'B}^{T_A} (\rho_A \otimes \frac{1}{d_B}\mathds{1}_B)U_{A'B}^{T_A\dag}\ket{\Gamma}_{AA'}\\
        =\Tr_A(\rho_A\otimes\frac{1}{d_B}\mathds{1}_B)=\frac{1}{d}\mathds{1}_B.
    \end{gather}
     Here, we used the property of $\ket{\Gamma}$ that $(\mathds{1}_{A'}\otimes O_A)\ket{\Gamma}=(O_{A'}^T\otimes\mathds{1}_A)\ket{\Gamma}$ for any operator $O$. Therefore $U$ is the catalysis unitary for a catalysis that uses $\frac{1}{d_B}\mathds{1}_B$ as the catalyst.
    
    Conversely, assume that $U$ is the catalysis unitary of a catalysis that uses an arbitrary quantum state $\sigma$ as its catalyst. From Proposition \ref{prop:comp}, we can assume that $\sigma=\frac{1}{d_B}\mathds{1}$. We input $\ket{\Gamma}_{AA'}$ into the catalysis. If we trace out the system $A$ after applying $U$ to $A$ and $B$, we get $U^{T_A}(\mathds{1}_{A'} \otimes \frac{1}{d_B}\mathds{1}_B)U^{T_A\dag}=\frac{1}{d_B} U^{T_A}U^{T_A\dag}$ for a similar reason with that of the previous case. However, this state should be $\Tr_A(\mathds{1}_{A'}\otimes U)\dyad{\Gamma}_{AA'}\otimes \frac{1}{d_B}\mathds{1}_B(\mathds{1}_{A'}\otimes U^\dag)=\mathds{1}_{A'}\otimes\frac{1}{d_B}\mathds{1}_B$ since the catalyst should remain unchanged regardless of the input state \cite{lie2019unconditionally}. This proves that $U^{T_A}$ is unitary.
\end{proofEnd}
The class of bipartite unitary operators with unitary partial transpose was previously known as the bipartite unitary operators that induce unital maps regardless of ancillary state \cite{deschamps2016some, benoist2017bipartite}. Theorem \ref{thm:catauni} adds an operational meaning to those bipartite unitary operators and we can see that only unital maps can be implemented through catalysis. We remark that, however, this characterization of catalysis unitary only applies to the case of implementation of quantum maps, not to the state transition between two specific quantum states.

On the Hilbert space associated a bipartite system, e.g. $\mathcal{H}_A\otimes \mathcal{H}_B$, we define the swapping operator $F:=\sum_{ij} \dyad{i}{j} \otimes \dyad{j}{i}.$ We remark that the partial transposes of $U^\dag$ and $FUF$ are also unitary operators. From Theorem \ref{thm:catauni}, we get the following Corollary.
\begin{corollary}
    A catalysis unitary operator $U$'s inverse $U^\dag$ and party-swapped version $FUF$ are also catalysis unitary operators.
\end{corollary}

Examining if a randomness source is compatible with a given catalysis unitary is seemingly complicated, but we show that actually there is an easy method of examining the compatibility. One need not examine the invariance of the randomness source for every input as it is enough to check the case of the maximally mixed input.

\begin{theoremEnd}{prop}
    A randomness source $\sigma$ is compatible with a catalysis unitary operator $U$ if and only if its von Neumann entropy is preserved for the maximally mixed input state, i.e.,
    \begin{equation}
        S\left(\Tr_A U\left(\frac{\mathds{1}_A}{d}\otimes\sigma_B\right)U^\dag\right)=S(\sigma_B).
    \end{equation}
\end{theoremEnd}
\begin{proofEnd}
    Consider the system $A$ is initially a part of a maximally entangled state $\ket{\Phi}_{RA}=\frac{1}{\sqrt{d}}\sum_{i=1}^d \ket{ii}_{RA}$ whose marginal state on $A$ is $\frac{1}{d}\mathds{1}_A$. The catalysis condition (\ref{eqn:catal}) is satisfied if and only if $RB$ is in a product state after applying $U$ to $AB$. Note that the mutual information $I(R:B)=S(R)+S(B)-S(RB)$ is zero if and only if the composite system $RB$ is in a product state. Since the system $R$ does not participate in the interaction, $R$ stays in the maximally mixed state, i.e. $S(R)=\log_2  d$. The composite system $RB$ is in $U^{T_A}(\frac{\mathds{1}_R}{d}\otimes\sigma_B)U^{T_A\dag}$ and $U_{RB}^{T_A}$ is also a unitary operator, hence $S(RB)=\log_2  d + S(\sigma)$. Therefore $I(R:B)=S(B)-S(\sigma)$ and $S(B)= S(\Tr_A U(\frac{\mathds{1}}{d}\otimes\sigma_B)U^\dag)$, we get the wanted result.
\end{proofEnd}

A special class of catalyses is \textit{classical catalysis}. In classical catalysis, the catalysis unitary is a controlled unitary operator which is conditioning on the eigenbasis of the catalyst. In other words, a classical catalysis is a \textit{random unitary operation} $\{U_X\}$, i.e. $\rho \mapsto \sum_x p_x U_x \rho U_x^\dag$, with the corresponding probability distribution $p_x=Pr(X=x)$.

In previous studies, advantages of quantum catalysts over classical catalysts have been discovered multiple times \cite{boes2018catalytic, lie2019unconditionally, lie2020uniform}. We introduce another specific functionality of quantum catalyst in the maximally mixed state in the following Theorem. Note that a unistochastic matrix is a doubly stochastic matrix which is the Schur square (component-wise square of absolute value) of a unitary matrix.

\begin{theoremEnd}{prop}
    A quantum catalyst in the $d$-dimensional maximally mixed state can be used to implement a random unitary operation $\{U_X\}$ followed by another random unitary operation $\{V_Y\}$, where $Pr(X=x)=Pr(Y=y)=\frac{1}{d}$ and $[U_x,V_y]=0$ for all $x$ and $y$, and the conditional probability matrix $Pr(Y=y|X=x)$ is unistochastic.
\end{theoremEnd}

A special class of unistochastic matrices is the family of stochastic matrices with uniform components. Such matrix is the Schur square of the discrete Fourier transform unitary matrix $F=(F_{nm})$, whose components are given as $F_{nm}=\frac{1}{\sqrt{d}}\exp(i2\pi nm/d)$.

\begin{corollary} \label{coro:double}
    A quantum catalyst in the $d$-dimensional maximally mixed state can be used to implement arbitrary two independent consecutive mutually commuting rank-$d$ random unitary operations.
\end{corollary}

Corollary \ref{coro:double} significantly generalizes the results of Ref. \cite{boes2018catalytic} and \cite{lie2020randomness, lie2020uniform}, thus strengthens the qualitative statement `quantum randomness is twice as strong as classical randomness'. It is still unclear, however, if the classical randomness source of double the size of quantum randomness can simulate the latter.

\subsection{Correlational aspect of randomness}\label{subsec:corasp}

 Catalysis of quantum randomness \cite{boes2018catalytic, lie2019unconditionally, lie2020randomness} was made possible by explicitly treating randomness sources as a quantum system. On the other hand, we intuitively know that randomness is consumed by building up correlation with the source of it. Therefore, we will generalize the explicit approach by explicitly treating the correlation with the randomness source as a bipartite quantum state.

 Suppose again that an agent $A$ is allowed to use a system $B$ called catalyst in the quantum state $\sigma_B$. $A$ is allowed to interact with $B$ but the system $B$ should stay in its original state $\sigma_B$ after every interaction. Let $\sigma_{A_1B}$ be the bipartite state shared by $A$ and $B$ throughout previous interactions. We will call such a bipartite state $\sigma_{A_1B}$ the \textit{intermediate} and its marginal state $\sigma_B$ the catalyst. (See FIG. \ref{fig:catalysis}.) Now, suppose that $A$ is trying to implement the state transition $\rho_{A_1} \otimes \sigma_{A_2} \mapsto \tau_{A_1A_2}$ for a new input state $\rho$ with some tripartite unitary operator $U$ in the following manner,

\begin{equation}
    \Tr_B U(\rho_{A_1} \otimes \sigma_{A_2B})U^\dag = \tau_{A_1A_2}.
\end{equation}
In addition to this, we require the catalysis constraint that $\sigma_{B}$ should be left unchanged, i.e.
\begin{equation}
    \Tr_A U(\rho_{A_1} \otimes \sigma_{A_2B})U^\dag = \sigma_{B}.
\end{equation}
Here, both systems $A_1$ and $A_2$ are collectively denoted as $A$. We let $\tau_{A_1A_2B}:=U(\rho_{A_1} \otimes \sigma_{A_2B})U^\dag$ and we will refer to this state as the output intermediate of the process. The following result shows that the mutual information of intermediate quantifies the amount of randomness already extracted from a catalyst.

\begin{theoremEnd}{theorem} \label{thm:main}
    The mutual information of the intermediate changes by the entropy production by the state transition, i.e.
    $I(A_1A_2:B)_\tau - I(A_2:B)_\sigma =  S(\tau_{A_1A_2})-S(\rho_{A_1}\otimes \sigma_{A_2})$.
\end{theoremEnd}
\begin{proof}
    We have
     $$I(A_1A_2:B)_\tau=S(A_1A_2)_\tau+S(B)_\tau-S(A_1A_2B)_\tau.$$
     Since $S(B)_\tau=S(B)_\sigma$ and $S(A_1A_2B)_\tau = S(\rho) + S(A_2B)_\sigma$ from the fact that unitary operators preserve the von Neumann entropy, we have
     \begin{equation} \label{eqn:I1}
         I(A_1A_2:B)_\tau = S(\tau_{A_1A_2}) -S(\rho) +  S(B)_\sigma - S(A_2B)_\sigma.
     \end{equation}
     On the other hand, we have
     $$I(A_2:B)_\sigma=S(A_2)_\sigma + S(B)_\sigma - S(A_2B)_\sigma.$$
     By subtracting the last equation from (\ref{eqn:I1}) and using the additivity of the von Neumann entropy, we get the wanted result.
\end{proof}
An important special case of such catalytic state transition is implementation of  a quantum map $\Phi$ independent of the products of previous interactions, i.e. 

\begin{equation}
      \Tr_B U(\rho_{A_1} \otimes \sigma_{A_2B})U^\dag = \Phi(\rho)_{A_1}\otimes\sigma_{A_2},
\end{equation}
and
\begin{equation}
      \Tr_{A} U(\rho_{A_1} \otimes \sigma_{A_2B})U^\dag = \sigma_{B},
\end{equation}
 for any input state $\rho$. In that case, the entropy production is exactly same with that by the quantum map $\Phi$, i.e. $I(A_1A_2:B)_\tau - I(A_2:B)_\sigma = S(\Phi(\rho))-S(\rho).$ When this is done, we will say that $\Phi$ is implemented catalytically with the intermediate $\sigma_{A_2B}$ and call $U$ as the generalized catalysis unitary operator.

 We remark that Theorem \ref{thm:main} opens up an unexplored application of randomness sources, namely their usage as a randomness absorbent. If the intermediate was initially given as a highly correlated state, then the source can be used to implement entropy-decreasing maps by decreasing the mutual information of the intermediate. This aspect of quantum catalyst will be discussed in Section  \ref{subsec:absorption}. However, the premise of this application is rather different; it requires the intermediate to be correlated in a \textit{known} form. Such a situation does not happen if the whole process has started from an uncorrelated catalyst and only accepts unknown input states. Note that consecutive implementation of quantum maps (say) $\Phi_1, \Phi_2, \dots$ is actually equivalent to a single implementation of tensor product of aforementioned quantum maps, i.e. $\Phi_1 \otimes \Phi_2 \otimes \dots$. Therefore, for that case, we can always assume every intermediate $\sigma_{AB}$ has the form $\sigma_{AB}=W(\rho_A\otimes \sigma_B)W^\dag$ for some catalysis unitary $W$. It leaves the study on catalysis with arbitrarily correlated catalysts as an open problem.

On the other hand, since a unital quantum map never decreases the entropy of its input state, i.e. $S(\Phi(\rho))-S(\rho) \geq 0,$ implementing a unital map only increases the mutual information of the intermediate. Unital maps are important since the class of unital maps coincides with the class of quantum maps that never decreases the entropy of its input state. It follows that randomness can be both deposited into and withdrawn from the intermediate.

Now, by maximizing the entropy production or reduction, we can lower bound the size of catalyst required for implementing a given map $\Phi$. Note that by implementing a quantum map $\Phi$, simultaneously one also implements a multipartite quantum map $\mathcal{I} \otimes \Phi$, where $\mathcal{I}$ could the identity map of the space of operators on an arbitrary Hilbert space. We can see that this result subsumes the previous results on the randomness costs for catalysis.

\begin{corollary} \label{coro:cost}
    The randomness source $\sigma_B$ for a catalysis should have the entropy no smaller than the maximal entropy variance of the target map $\Phi$, i.e. $S(\sigma_B)\geq \frac{1}{2} \max_\rho |S((\mathcal{I}\otimes \Phi)(\rho))-S(\rho)|.$ If $B$ is a classical system, we have $S(\sigma_B)\geq \max_\rho |S((\mathcal{I}\otimes \Phi)(\rho))-S(\rho)|.$
\end{corollary}

\begin{proofEnd}
    We consider the case where $\mathcal{I}\otimes \Phi$ is implemented with an input state $\rho$ From the Araki-Lieb inequality \cite{araki1970entropy} and Theorem \ref{thm:main}, we have 
    \begin{align*}
        S(B)_\sigma =& S(B)_\tau\\
        \geq& r_{C,Q} \max\{ I(A_1A_2:B)_\tau ,I(A_2:B)_\sigma \} \\
        \geq& r_{C,Q} |I(A_1A_2:B)_\tau - I(A_2:B)_\sigma|\\
        =& r_{C,Q} |S((\mathcal{I}\otimes \Phi)(\rho))-S(\rho)|,
    \end{align*}
     where $r_{C}=1$ for classical catalyst and $r_Q=\frac{1}{2}$ for quantum catalyst. Since the inequality holds for arbitrary input $\rho$, by maximizing over $\rho$, we get the desired result. 
\end{proofEnd}
    We will call $S(\Phi):=\max_\rho \left[S(\Phi(\rho))-S(\rho)\right]$ the maximal local entropy production of $\Phi$ and $S^G(\Phi):=\max_\rho \left[S((\mathcal{I}\otimes\Phi)(\rho))-S(\rho)\right]$ the maximal global entropy production of $\Phi$. We similarly define their R\'{e}nyi entropy counterparts, $S_\alpha(\Phi)$ and $_\alpha^G(\Phi)$, in a similar way. Note that $S_\alpha^G\geq S_\alpha$ (See Section \ref{subsec:absorption}).
    
    Note that the maximal entropy production always can be achieved with a pure state input. This can be shown from noting that for a general bipartite input state $\rho_{AB}$, there exists a purifying system $E$ so that $\rho_{ABE}$ is a pure state and the entropy production by $\Phi_A$ is given by $S(AB)_\tau-S(E)_\tau$ where $\tau_{ABE}=(\Phi_A \otimes \mathcal{I}_{BE})(\rho_{ABE})$. By using the Araki-Lieb inequality \cite{araki1970entropy}, we get $S(AB)_\tau-S(E)_\tau \leq S(ABE)_\tau$ where $S(ABE)_\tau$ can be also interpreted as the entropy production by $\Phi_A$ for the bipartite pure state input $\rho_{ABE}$.
    
     For example, for the dephasing map $\mathcal{D}$ with respect to the computational basis, by choosing a pure state $\rho=\dyad{+}$ with $\ket{+}=\frac{1}{\sqrt{d}}\sum_i \ket{i}$, we have $\mathcal{D}(\dyad{+})=\frac{1}{d}\mathds{1}$, thus the maximal entropy production is achieved, i.e. $ S(\mathcal{D}(\dyad{+}))-S(\dyad{+})=\log_2  d$. For the erasure map $\mathcal{E}(\rho):=\frac{1}{d}\mathds{1}$, by choosing $\Phi=\mathcal{I}\otimes\mathcal{E}$ and the input state $\rho=\dyad{\Psi}$ with an arbitrary maximally entangled state $\ket{\Psi}$ (e.g. $\ket{\Psi}=\frac{1}{\sqrt{d}}\sum_i \ket{ii}$), we get $ S((\mathcal{I}\otimes\mathcal{D})(\dyad{\Psi}))-S(\dyad{\Psi})  = 2\log_2  d$.

\subsection{Catalytic entropies} \label{subsec:catent}

\begin{figure}[t]
    \includegraphics[width=.5\textwidth]{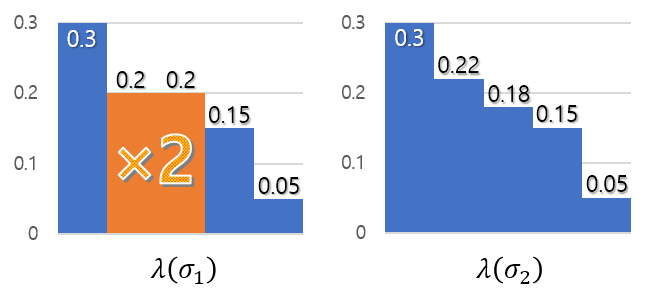}\label{fig:entropy}
    \caption{Spectrums of two density matrices. Each probability $p_i$ contributes to the catalytic entropy by $-p_i\log_2  p_i$, however, if there is degeneracy, then the same contributes by the double, i.e. $-2p_i\log_2p_i$. Although their von Neumann entropies are very close, i.e.$|S(\sigma_1)-S(\sigma_2)|<0.004$, their catalytic entropies differ by almost 1 bit.}
\end{figure}

A naturally following question is how to measure the remaining free randomness in a catalyst. Theorem \ref{thm:main} shows that the mutual information of the intermediate is the total amount of entropy extracted from a catalyst. Hence, if we can identify the maximal entropy extractable from a catalyst, then, by subtracting the mutual information of the intermediate from it, we can calculate the amount of free randomness in the catalyst.
 
Therefore, we completely characterize the amount of entropy extractable from an arbitrary quantum catalyst. By the eigenspace decomposition of a quantum state $\sigma$ we mean the decomposition of the form $\sigma=\sum_i \lambda_i \Pi_i$ with $\{\lambda_i\}$ being the eigenvalues of $\sigma$ and $\Pi_i$ being the orthogonal projector onto the eigenspace corresponding to $\lambda_i$ such that $\Pi_i\Pi_j=0$ if $\lambda_i\neq \lambda_j$. It was shown in Ref. \cite{lie2020uniform} that uniformness or degeneracy of eigenvalues of a quantum state boosts its capability as a catalytic randomness source, so we define the average degeneracy $\Delta(\sigma)$ of quantum state $\sigma$ counted in bits as $\Delta(\sigma) := \sum_i \lambda_i r_i \log_2  r_i$. For example, $\Delta(\sigma)$ is zero for a non-degenerate $\sigma$ and $\Delta(\sigma)$ achieves its maximal value, $S(\sigma)$, when $\sigma$ is completely uniform.

In the following theorem, we show that, in addition to the von Neumann entropy of the catalyst, the average degeneracy acts as the bonus extractable entropy of the catalyst.
 
\begin{theoremEnd}{theorem} \label{thm:character}
    For arbitrary randomness source $\sigma$ with the eigenspace decomposition $\sigma=\sum_i \lambda_i \Pi_i$, the maximal entropy production from $\sigma$ is $S(\sigma) + \Delta(\sigma)$.
\end{theoremEnd}

\begin{proofEnd}
    Consider a catalytic map $\Phi$ using $\sigma$ as a catalyst given as
    \begin{equation}
        \Phi(\rho)=\Tr_B U(\rho_A \otimes \sigma_B)U^\dag.
    \end{equation}
    It follows that $\left[U,\mathds{1}_A \otimes \sigma_B \right]=0$ \cite{lie2020uniform}. Subsequently, $U$ also commutes with every projector onto eigenspace of $\sigma$, i.e. $\left[U,\mathds{1}_A \otimes \Pi_i \right]=0$ for all $i$. Therefore, by letting $U_i := (\mathds{1}_A \otimes \Pi_i) U (\mathds{1}_A \otimes \Pi_i)$, we can see that each $U_i$ is a unitary operator on $\text{supp}(\mathds{1}_A \otimes \Pi_i)$. It allows us to decompose $\Phi$ into the following form,
    \begin{equation}\label{eqn:convex}
        \Phi(\rho)=\sum_i  \lambda_i r_i \Tr_B U_i(\rho_A \otimes \pi_i)U_i^\dag.
    \end{equation}
    Where $\pi_i := \frac{1}{r_i}\Pi_i$. Thus $\Phi$ can be considered a probabilistic mixture of subchannels, i.e. $\Phi=\sum_i \lambda_i r_i \Phi_i$ where $\Phi_i (\rho) := \Tr_B U_i(\rho_A \otimes \pi_i)U_i^\dag$. Note that $\sum_i \lambda_i r_i = 1$. Since each $\Phi_i$ is a catalysis using a uniform catalyst, each of them can produce entropy up to $2\log_2  r_i$. Now, for arbitrary pure state input $\phi$, the entropy production by $\Phi$ is given by $S((\mathcal{I}\otimes \Phi)(\phi))$, which is upper bounded by $H(\lambda_i r_i) + \sum_i \lambda_i r_i S((\mathcal{I}\otimes \Phi_i)(\phi)).$ The latter terms is, in turn, upper bounded by 2$\sum_i \lambda_i r_i \log_2  r_i$. Therefore we get the upper bound $H(\lambda_i r_i) + 2\sum_i \lambda_i r_i \log_2  r_i = S(\sigma) + \sum_i \lambda_i r_i \log_2  r_i$.
    
    We will show that this upper bound is indeed achievable. First we let $n$ be the number of different eigenvalues of $\sigma$ and $R$ be the least common multiple of all $r_i^2$. Suppose that the system $A$ is composed of two systems, $n$-dimensional $A_1$ and $R$-dimensional $A_2$. Similarly we consider their reference systems $E_1$ and $E_2$ with the same dimensions. We define the following entangled state
    $$\ket{\Psi}_{AE}=\frac{1}{\sqrt{nR}}\sum_{i=1}^n\sum_{j=1}^{R}\ket{ij}_{A_1A_2}\otimes \ket{ij}_{E_1E_2}.$$
    Next, consider the following unitary operator $U$ acting on $A$ and $B$.
    \begin{equation}
        U=\sum_{i=1}^n \sum_{j=1}^{r_i^2} V_{A_1i} \otimes P_{A_2j}^{(i)}  \otimes W_{Bj}^{(i)}.
    \end{equation}
    Here, $P_j^{(i)}$ are mutually orthogonal projectors satisfying $\Tr P_j^{(i)} = R/r_i^2$ on $A_2$ satisfying $\sum_j P_j^{(i)} = \mathds{1}_{A_2}$ for all $i$. Also, $\{V_m\}$ and $\{W_m^{(i)}\}$ are the sets of orthogonal unitary operators on respectively $A_1$ and supp$\Pi_i$ satisfying $\Pi_i W_m^{(i)} \Pi_i = W_m^{(i)}$. One can check that $U^\dag U = U U^\dag = \mathds{1}_{AB}$. The catalytic map $\Phi$ defined in such a way increases the entropy of the pure input state $\Psi_{AE}$ by $H(\lambda_i r_i) + 2\sum_i \lambda_i r_i \log_2  r_i=S(\sigma) + \sum_i \lambda_i r_i \log_2  r_i$.
\end{proofEnd}
This type of relation between the non-degeneracy and the entropy of quantum state can be extended to the min-entropy. Note that the maximal extractable von Neumann entropy of a quantum state $\sigma$ with the eigenspace decomposition $\sigma=\sum_i \lambda_i \Pi_i$, $S(\sigma)+\Delta(\sigma)$ can be we written as 
\begin{equation}
   S^\diamond(\sigma):= -\sum_i \lambda_i r_i \log_2  (\lambda_i/r_i),
\end{equation}
which we will call the \textit{catalytic (von Neumann) entropy} of the catalyst $\sigma$.  This is the average of quantities $-\log_2  (\lambda_i/r_i)$, which can be interpreted as the `catalytic power' of each sub block $\Pi_i$ in the catalyst $\sigma$. Therefore, its natural `min-entropy'-like generalization would be 
\begin{equation}
    S_{\min{}}^{\diamond}(\sigma):=-\max_i \log_2  (\lambda_i/r_i),
\end{equation}
which we will call the \textit{catalytic min-entropy} of $\sigma$. We remark that the min-entropy cannot exceed the catalytic min-entropy. Also, just as ordinary quantum R\'{e}nyi entropies, we have the order relation $S_{\min{}}^\diamond\leq S^\diamond.$

In the following Theorem, we will show that this catalytic min-entropy is indeed the maximal min-entropy extractable from a given catalyst.

\begin{theoremEnd}{theorem} \label{thm:tradeoff2}
    For arbitrary randomness source $\sigma$, the maximal extractable min-entropy from $\sigma$ is the catalytic min-entropy of $\sigma$, $S_{\min{}}^\diamond(\sigma)$.
\end{theoremEnd}

\begin{proofEnd}
    Consider an arbitrary catalysis $\Phi$ whose catalyst is $\sigma$. We employ the same decomposition of $\Phi=\sum_i \lambda_i r_i \Phi_i$ in the proof of Theorem \ref{thm:character}. The following Lemma will be helpful for the proof.
    \begin{theoremEnd}{lemma} \label{lem:minineq}
    Let a quantum state $\rho$ be a convex sum of other quantum states, i.e. $\rho=\sum_i p_i \rho_i.$ Then we have
    $$S_{\min{}}(\rho) -S_{\min{}}(\rho_i) \leq - \log_2  p_i,$$
    for every $i.$
    \end{theoremEnd}
    It follows from the facts that $2^{-S_{\min{}}(\rho)}=\max_{\ket{\psi}} \bra{\psi}\rho\ket{\psi}$ and that, for $\ket{\phi}$ such that $2^{-S_{\min{}}(\rho_i)}=\bra{\phi}\rho_i\ket{\phi}$, $p_i 2^{-S_{\min{}}(\rho_i)}\leq \sum_i p_i \bra{\phi}\rho_i\ket{\phi} = \bra{\phi} \rho \ket{\phi} \leq \max_{\ket{\psi}}\bra{\psi}\rho\ket{\psi}=2^{-S_{\min{}}(\rho)}$.
    
    For arbitrary bipartite state $\phi$, we apply this Lemma by substituting $\rho=(\mathcal{I}\otimes \Phi)(\phi)$, $\rho_i=(\mathcal{I}\otimes \Phi_i)(\phi)$ and $p_i = \lambda_i r_i$. Now, as each $\Phi_i$ is a catalysis with the corresponding catalyst $\pi_i$, from the weak subadditivity of R\'{e}nyi entropy \cite{van2002renyi}, we have
    $$S_{\min{}}((\mathcal{I}\otimes \Phi_i)(\phi))-S_{\max{}}(\pi_i)\leq S_{\min{}}(\pi_i).$$
    However, since the catalyst $\pi_i$ is uniform, we have $S_{\min{}}(\pi_i)=S_{\max{}}(\pi_i)=\log_2  r_i$, thus an upper bound $S_{\min{}}((\mathcal{I}\otimes \Phi_i)(\phi)) \leq 2 \log_2  r_i$ follows. Combining all the results, we have
    $$S_{\min{}}((\mathcal{I}\otimes \Phi)(\phi)) \leq  - \log_2  (\lambda_i/r_i).$$
    This result holds for every $i$ and pure state $\phi$, so we get
    \begin{equation}
        \max_\phi S_{\min{}}((\mathcal{I}\otimes \Phi)(\phi)) \leq  - \max_i \log_2  (\lambda_i/r_i),
    \end{equation}
    where the left hand side can be interpreted as the maximal min-entropy production on pure states by $\Phi$. We claim that, from Lemma \ref{lem:minineq}, it follows that actually the maximal min-entropy production can be achieved with a pure state input. It can be shown by substituting $\rho=(\mathcal{I}\otimes\Phi)(\tau)$, where $\tau$ is an arbitrary (possibly mixed) input state, and $\rho_i=(\mathcal{I}\otimes\Phi)(\tau_i)$, where $\tau=\sum_i t_i \tau_i$ is the spectral decomposition of $\tau$ so that each $\tau_i$ is a pure eigenstate of $\tau$ corresponding to the eigenvalue $t_i$ and $p_i=t_i$. By picking the index $k$ such that $t_k=2^{-S_{\min{}}(\tau)}$ and using the fact that $S_{\min{}}((\mathcal{I}\otimes\Phi)(\tau_k))\leq \max_\phi S_{\min{}}((\mathcal{I}\otimes\Phi)(\phi)) $ where the maximization is over every pure state $\phi$ so that the right hand side is the maximal min-entropy production of $\Phi$ on pure state inputs, we get the wanted result.
    
    Conversely, the same $\ket{\Psi}$ and $U$ of the proof of Theorem \ref{thm:character} achieves the maximal min-entropy extraction of $-\max_i \log_2  (\lambda_i/r_i)$ as the spectrum of the output state of the process is $\{\lambda_i/r_i\}$.
\end{proofEnd}
In a similar way, we can define the \textit{catalytic R\'{e}nyi entropy} $S_\alpha^\diamond$ for every $\alpha \in (0,1)\cup(1,\infty)$ as
\begin{equation} 
    S_\alpha^\diamond(\sigma):=\frac{1}{1-\alpha}\log_2 \sum_i \lambda_i^\alpha r_i^{2-\alpha}.
\end{equation}
Similarly to the catalytic min-entropy, we can also define the catalytic max-entropy $S_{\max{}}^\diamond(\sigma):=\log_2  \sum_i r_i^2.$ Then,we have the order relation $S_{\min{}}^\diamond\leq S_{\alpha}^\diamond \leq S_\beta^\diamond \leq S_{\max{}}^\diamond$ for $0<\beta<\alpha$.  Like the both previously defined catalytic entropies, catalytic R\'{e}nyi entropy also characterizes the maximally extractable R\'{e}nyi entropy with the corresponding $\alpha$.
\begin{theoremEnd}{theorem} \label{thm:traderen}
    For arbitrary randomness source $\sigma$, the maximal extractable R\'{e}nyi entropy from $\sigma$ is the catalytic R\'{e}nyi entropy of $\sigma$, $S_{\alpha}^\diamond(\sigma)$.
\end{theoremEnd}
\begin{proofEnd}
    The proof is basically identical with that of Theorem \ref{thm:character}, except that we use the  facts \cite{nielsen2000probability} that 
    \begin{align}
        (\mathcal{I}\otimes \Phi)(\phi) =& \sum_i \lambda_i r_i (\mathcal{I}\otimes \Phi_i)(\phi)\\
        \succ& \bigoplus_i \lambda_i r_i (\mathcal{I}\otimes \Phi_i)(\phi),
    \end{align}
     and that for each $i$, $(\mathcal{I}\otimes \Phi_i)(\phi) \succ \frac{1}{r_i^2} \mathds{1}_{r_i^2}$ where $\mathds{1}_{r_i^2}$ is a projector with rank $r_i^2$. Here, $\oplus$ operation is the direct sum operation which can be interpreted in terms of tensor product as $\bigoplus_i O_i = \sum_i \dyad{i}\otimes O_i$ for a set of operators $\{O_i\}$ with an orthonormal basis $\{\ket{i}\}$. The latter majorization relation follows from the fact that the rank of each $(\mathcal{I}\otimes \Phi_i)(\phi)$ is upper bounded by $r_i^2$ from the triangular inequality of the max-entropy. From the Schur concavity of R\'{e}nyi entropy, we have $S_\alpha((\mathcal{I}\otimes \Phi)(\phi))\leq S_\alpha(\bigoplus_i \lambda_i r_i^{-1}\mathds{1}_{r_i^2})=\frac{1}{1-\alpha} \log_2  \sum_i \lambda_i^\alpha r_i^{2-\alpha}.$ Again, the maximal entropy extraction is achievable with pure states since for any mixed state input $\rho$ with the spectral decomposition $\rho=\sum_i a_i \phi_i$, we have 
    \begin{gather}
        (\mathcal{I}\otimes \Phi)(\rho) =\sum_i a_i (\mathcal{I}\otimes \Phi)(\phi_i)\\
        \succ \bigoplus_i a_i (\mathcal{I}\otimes \Phi)(\phi_i) \succ \bigoplus_i a_i \left(\bigoplus_j \lambda_jr_j^{-1}\mathds{1}_{r_j^2} \right)\\
        = \sum_i a_i \dyad{i} \otimes \left(\bigoplus_j \lambda_jr_j^{-1}\mathds{1}_{r_j^2} \right).
    \end{gather}
    Note that $S_\alpha(\sum_i a_i\dyad{i}) = S_\alpha(\rho)$. Repeatedly, from the Schur concavity of $S_\alpha$, it follows that
    $S_\alpha((\mathcal{I}\otimes \Phi)(\rho))\leq S_\alpha \left(\left(\sum_i a_i \dyad{i} \right)\otimes \left(\bigoplus_j \lambda_j r_j^{-1}\mathds{1}_{r_j^2} \right) \right)=S_\alpha(\rho)+S_\alpha^\diamond(\sigma)$, i.e. the R\'{e}nyi entropy production by $\Phi$ on $\rho$, $S_\alpha((\mathcal{I}\otimes \Phi)(\rho)) -  S_\alpha (\rho)$ is upper bounded by $S_\alpha^\diamond(\sigma).$

    Conversely, this bound can be achieved with the same example in the proof of Theorem \ref{thm:character}.
\end{proofEnd}
Since $\lim_{\alpha \to 1}S_\alpha^\diamond=S^\diamond$ and $\lim_{\alpha \to \infty}S_\alpha^\diamond=S_{\min{}}^\diamond,$ Theorem \ref{thm:traderen} subsumes Theorem \ref{thm:character} and \ref{thm:tradeoff2} but we gave different proofs using properties specific for each entropic quantity.

In previous works, it was shown that quantum maps that erase more information require more randomness resources \cite{lie2020randomness,lie2020uniform}. We show here that the same relation holds for the catalytic R\'{e}nyi entropies.

\begin{theoremEnd}{corollary} \label{coro:tradeoff}
For a $d$-dimensional catalytic map $\Phi$ with the entanglement-assisted classical capacity $C_{EA}(\Phi)$ utilizing a catalyst $\sigma$, the following inequality holds.
    \begin{equation} \label{eqn:tradeoff}
        2\log_2  d - C_{EA}(\Phi) \leq S_{\min{}}^\diamond(\sigma).
    \end{equation}
\end{theoremEnd}
\begin{proofEnd}
    Consider the decomposition of $\Phi$ of the from of (\ref{eqn:convex}), which we re-express as $\Phi=\sum_i \lambda_i r_i \Phi_i$. By denoting the entanglement-assisted classical capacity of $\Phi_i$ by $C_i$, we have the following inequality \cite{lie2020randomness}.
    \begin{equation}
        C_i-C_{EA}(\Phi) \leq - \log_2 \lambda_i r_i.
    \end{equation}
    However, from the proof of Theorem \ref{thm:character}, it follows that each $C_i$ is $d$-dimensional catalysis utilizing the catalyst $\pi_i$, we have the following inequality \cite{lie2020uniform}
    \begin{equation}
        2(\log_2  d - \log_2  r_i) \leq C_i. 
    \end{equation}
    From these two inequalities we get the following relation.
    \begin{equation}
        2\log_2  d - C_{EA}(\Phi) \leq - \log_2  (\lambda_i/r_i).
    \end{equation}
    By maximizing $\log_2  r_i$ over $i$ we get $2\log_2  d - C_{EA}(\Phi) \leq \Delta_{\max{}}(\sigma) - \log_2  \lambda_i$. As it holds for every $i$, we get the wanted result.
\end{proofEnd}

\subsection{Catalysis under conservation law} \label{subsec:catund}
   
    From the proof structure of the previous results, the relation between the degeneracy and the entropy of catalyst follows from the relation between the decomposability of the given catalysis into sub-catalyses and the entropy of catalyst. For example, every classical catalyst can be decomposed into rank-1 catalysts, therefore requires more entropy to implement the same quantum map.
 
To be more precise, even when we decompose a given catalyst $\sigma$ more finely so that its eigenspace decomposition $\sigma=\sum_i \lambda_i \Pi_i$ need not have distinct eigenvalues for different $i$'s, but still different $\frac{1}{r_i}\Pi_i$ are required to be orthogonal to each other and to be catalysts themselves for the same catalysis unitary, Theorem \ref{thm:character}, \ref{thm:tradeoff2} and Corollary \ref{coro:tradeoff} still hold. For instance, even when a catalyst is maximally degenerate, i.e. $\sigma=\frac{1}{d}\sum_i \dyad{i}$, if each projector $\dyad{i}$ is preserved when used instead of $\sigma$ itself for the same catalysis process, then $r_i=1$ for every $i$ so that $S_\alpha(\sigma)=S_\alpha^\diamond(\sigma)$. In this case, we assume that entropies $S_\alpha^\diamond$ depending on $\{r_i\}$ themselves depend on the catalysis and say the catalysis or the catalyst has degeneracy when $r_i>1$ for some $i$. When its catalyst has a eigenspace decomposition $\sum_i \lambda_i \Pi_i$ with the aforementioned property, we will call the vector $(r_1,\cdots,r_n)$ the degeneracy vector of the given catalysis or catalyst.

 Sometimes there exist limits on the level of degeneracy without complete specification of the form of catalysis. A naturally occurring case is where a conservation law is imposed on catalyst system. More precisely, consider the physical quantity $Q$ associated with a projective measurement $\{\Pi_{q}\}_{q=1}^n$ on the catalyst system such that $\sum_q \Pi_q = \mathds{1}$. It implies that any quantum state or unitary operation on this system commutes with every projector $\Pi_q$. Again, if we let $r_q=\Tr \Pi_q$, any catalysis unitary should have $(r_1,\cdots,r_n)$ as its degeneracy vector. Note that $\sum_i r_i = \rank \sigma$ when $\sigma$ is a would-be catalyst. One of the most canonical examples is catalysis through thermal operations using the heat bath in a thermal state as a catalyst.

 Let $\|\textbf{r}\|_2:=\sqrt{r_1^2 + \cdots + r_n^2}$ and let $t=(t_i)$ be the probability distribution formed by normalizing the squared degeneracy vector $\textbf{r}$, i.e. $t_i:=\norm{\textbf{r}}_2^{-2}r_i^2$. Then, we have the following expression for the catalytic R\'{e}nyi entropy of $\sigma$ in terms of R\'{e}nyi divergence.
 \begin{equation}
     S_\alpha^\diamond(\sigma)= 2\log_2  \|\textbf{r}\|_2 - D_\alpha \barpair*{\lambda_i r_i}{t_i}.
 \end{equation}
Here, $D_\alpha(p\|q):=\frac{1}{\alpha-1}\log_2 \sum_ip_i^\alpha q_i^{1-\alpha}$ is the R\'{e}nyi divergence between two probability distributions $p=(p_i)$ and $q=(q_i)$, which is nonnegative and is zero if and only if $p=q$. From this expression we get that the maximal catalytic R\'{e}nyi entropy can be achieved when $\lambda_i=\norm{\textbf{r}}_2^{-2}r_i$.
 
\begin{corollary} \label{coro:vector}
    For a catalysis with degeneracy vector $\textbf{r}=(r_1,\cdots,r_n)$, the maximal catalytic R\'{e}nyi entropy of compatible catalyst is $2\log_2 \norm{\textbf{r}}_2$.
\end{corollary}
 
 The catalyst achieving the maximal catalytic entropies in Corollary \ref{coro:vector} has the same catalytic entropies with the maximally uniform quantum catalyst with rank $\norm{\textbf{r}}_2$. Therefore, one can interpret that $\norm{\textbf{r}}_2$ is the effective dimension of a quantum catalyst under the restriction that degeneracy vector should be $\textbf{r}.$ Moreover, it is indeed possible to implement $\norm{\textbf{r}}_2^2$-dimensional dephasing map.

\begin{theoremEnd}{theorem}
   With a quantum catalyst with degeneracy vector $\textbf{r}=(r_1,\cdots,r_n)$, the $\norm{\textbf{r}}_2^2$-dimensional dephasing map can be implemented.
\end{theoremEnd}
\begin{proofEnd}
    We assume that the catalyst $\sigma$ has the eigenspace decomposition $\sigma=\norm{\textbf{r}}_2^{-1}\sum_m r_m \Pi_m$ with $\Tr \Pi_m = r_m$. Let $S_m:=\sum_{k=1}^{m-1}r_k^2$ with $S_1:=0$ and $\norm{\textbf{r}}_2^2\otimes r_m$-dimensional unitary operator $W_m$ be defined as $W_m := r_m^{-1/2} \sum_{i,j=1}^{r_m} \omega_m^{ij} Z^{S_m+ir_m+j}\otimes \dyad{m_i}{m_j}$, where $\omega_m$ is the $m$th root of unity and $\{\ket{m_i}\}$ is an orthonormal basis of the support of $\Pi_m$. Note that each $W_m$ is a catalysis unitary operator for the catalyst $r_m^{-1}\Pi_m$ that implements the random unitary map $\Phi_m(\rho):= r_m^{-2}\sum_{k=1}^{r_m^2} Z^{S_m+k}\rho Z^{-S_m-k}$. Then $\sum_m W_m$ is a catalysis unitary operator on $\norm{\textbf{r}}_2^2 \otimes \norm{\textbf{r}}_2$-dimensional space that implements a convex sum of $\Phi_m$, i.e. $\Phi(\rho)=\norm{\textbf{r}}_2^{-2}\sum_m r_m^2 \Phi_m(\rho)=\norm{\textbf{r}}_2^{-2}\sum_{k=1}^{\norm{\textbf{r}}_2^2} Z^k\rho Z^{-k}$, which is the $\norm{\textbf{r}}_2^2$-dimensional dephasing map with respect to the eigenbasis of $Z$.
\end{proofEnd}
These observations show that the notions `the maximally mixed state' and `the state providing maximal entropy' are no longer identical under the superselection rule. For example, for an electron in atom whose azimuthal quantum number is $l$ and magnetic quantum number $m$ with restriction $l\leq l_M$, if there is a superselection rule that forbids the superposition between states with different azimuthal quantum numbers, then the state that exhibits the maximal catalytic entropy is not the maximally mixed state 
\begin{equation}
    \frac{1}{(l_M+1)^2}\sum_{l=0}^{l_M}\sum_{m=-l}^l \dyad{l,m},
\end{equation}
but the state with the specific mixing probability
\begin{equation}
    \sum_{l=0}^{l_M}\frac{3(2l+1)}{(l_M+1)(2l_M+1)(2l_M+3)}\sum_{m=-l}^l \dyad{l,m},
\end{equation}
whose catalytic entropy is $\log_2[(l_M+1)(2l_M+1)(2l_M+3)/3]$.
For the case where the catalyst is a thermal state, i.e. $\sigma=e^{-\beta H}/Z$ with some Hamiltonian $H$, the energy levels $\{E_i\}$ should have the form $E_i = E_\infty -\log_2  r_i $ with some constant energy cap $E_\infty$.

\subsection{Depletion of catalyst} \label{subsec:depcat}

In this section, we will show that a randomness source can be actually \textit{depleted}. Suppose that, for a given catalyst $\sigma$, the maximal entropy production of a unital map $\Phi_1$ is already $S^\diamond(\sigma)$, i.e. $S(\Psi_1)=S^\diamond(\Psi_1)$. Can we catalytically implement another unital map $\Psi_2$ after implementing $\Psi_1$, or in other words, can we implement $\Psi_1\otimes\Psi_2$, with the catalyst $\sigma$? We answer this question negatively by proving the following result.

\begin{theoremEnd}{theorem} \label{thm:saturation}
    Consider catalysis processes with the catalyst $\sigma_B$ and let $\Psi_1$ and $\Psi_2$ be unital maps acting on $A_1$ and $A_2$ respectively. For arbitrary catalytical implementation of a quantum map $\Psi$ on $A_1A_2$ utilizing $\sigma_B$ such that $\Tr_{A_2}\circ\Psi = \Psi_1$ and $\Tr_{A_1}\circ\Psi=\Psi_2$, we have $\max_{\rho_1,\rho_2} I(A_1:A_2)_{\Psi(\rho_1\otimes\rho_2)}\geq S(\Psi_1)+S(\Psi_2)-S^\diamond(\sigma)$.  
\end{theoremEnd}
\begin{proofEnd}
    We first let $\rho_i$ be a quantum state that achieves $S(\Psi_i)=S(\Psi_i(\rho_i))-S(\rho_i)$ for $i=1,2$ and let $\Delta \mathcal{S}=S(\Psi_1)+S(\Psi_2)-S^\diamond(\sigma)$. Then, we get, omitting the subscript, i.e. $I(A_1:A_2)=I(A_1:A_2)_{\Psi(\rho_1\otimes\rho_2)}$,
    \begin{align*}
        I(A_1&:A_2)=S(\Psi_1(\rho_1))+S(\Psi_2(\rho_2))-S(\Psi(\rho_1\otimes\rho_2))\\
        &=S(\Psi_1)+S(\Psi_2)+S(\rho_1\otimes\rho_2)-S(\Psi(\rho_1\otimes\rho_2))\\
        &= \Delta \mathcal{S} + S^\diamond(\sigma)+S(\rho_1 \otimes \rho_2)-S(\Psi(\rho_1\otimes\rho_2))\\
        &\geq\Delta \mathcal{S}.
    \end{align*}
    Where the second equality holds since $S(\Psi_i)=S(\Psi_i(\rho_i))-S(\rho_i)$ for $i=1,2$ and  $S(\rho_1\otimes\rho_2)=S(\rho_1)+S(\rho_2)$, and the third inequality holds since $\Delta \mathcal{S}=S(\Psi_1)+S(\Psi_2)-S^\diamond(\sigma)$. The inequality holds since $S^\diamond(\sigma)$ is the maximally extractable entropy from $\sigma$ through catalysis and $\Psi$ itself is also being implemented catalytically, therefore $S^\diamond(\sigma)\geq S(\Psi(\rho_1\otimes \rho_2))-S(\rho_1\otimes \rho_2)$.
\end{proofEnd}
Theorem \ref{thm:saturation} implies that $\Psi_1\otimes\Psi_2$ cannot be implemented catalytically if the sum of their maximal entropy productions exceeds the catalytic entropy of the catalysis since $\Psi_1\otimes\Psi_2(\rho_1\otimes \rho_2)=\Psi_1(\rho_2)\otimes \Psi_2(\rho_2)$ is a product state for arbitrary $\rho_1$ and $\rho_2$ so its mutual entropy should be zero, but Theorem \ref{thm:saturation} forbids it. By substituting $\Psi_i \mapsto \mathcal{I}\otimes\Psi_i$ for $i=1,2$ in Theorem \ref{thm:saturation}, a useful Corollary follows.

\begin{corollary}
    For a pair of unital maps $\Psi_1$ and $\Psi_2$ such that $S^G(\Psi_1)+S^G(\Psi_2)>S^\diamond(\sigma)$, $\Psi_1\otimes \Psi_2$ cannot be implemented catalytically with the catalyst $\sigma$.
\end{corollary}

We remark that the requirement (\ref{eqn:catal}) is not actually requiring the state $\sigma_B$ to be used indefinitely by a single user, in the sense that it does not get deteriorated after each use for the agent consecutively using the randomness source. Using randomness source can be compared to checking a book out of a library. If a reader checks out the same book multiple times because she cannot finish the book in one read, then whenever she returns the book, it should be made sure that the book is in its original state, undamaged and unspoiled. Nonetheless, as an information resource, a book can be `depleted' to a particular reader when the reader finishes reading. As long as the book itself is maintained perfectly, however, the book can be read again and again by different readers. The book is a `catalyst' in this sense.

\section{discussion} \label{sec:discussion}

\subsection{Randomness absorption of correlated catalyst} \label{subsec:absorption}
The extremal case of entropy-decreasing map is the initialization map, which maps every input state to a single pure state. An initialization map cannot be implemented with a catalyst that is uncorrelated with the system, however, it is possible with an initially correlated intermediate. The observation that the amount of randomness required to decouple a correlated state effectively measures the correlation within it was made in Ref. \cite{groisman2005quantum}. Catalytic utilization of correlated intermediate can be understood as a converse of that observation. Nonetheless, surprisingly, a more correlated intermediate is not always more useful for catalytic implementation of initialization map. In fact, a highly restrictive form is required as the following Proposition shows.

\begin{theoremEnd}{prop} \label{prop:initial}
    A $d$-dimensional quantum catalyst compatible for implementation of a $d$-dimensional initialization map should be in the maximally mixed state and be part of an intermediate with the mutual information $\log_2  d$.
\end{theoremEnd}
\begin{proofEnd}
    We can assume that the target map $\Phi$ is given as $\Phi(\rho)=\dyad{0}$ without loss of generality. The maximal entropy decrease by $\Phi$ is $\log_2  d$ which can be achieved only with the maximally mixed input state $\frac{1}{d}\mathds{1},$ and the maximal entropy increase by $\mathcal{I}\otimes\Phi$ is also $\log_2  d$, achieved with a maximally entangled pure input state, e.g. $\frac{1}{d}\dyad{\Gamma}$.
    
    Therefore, the mutual information of the intermediate should be able to change by $\log_2  d$ in both directions. However, the mutual information of an intermediate for a $d$-dimensional catalyst is upper bounded by $2 \log_2  d$, which can only be achieved with the maximally mixed catalyst. It leaves $\log_2  d$ as the only possible value for the mutual information of the initial intermediate.
\end{proofEnd}
One example of such an intermediate is, when $d$ is an odd number, a $d$-dimensional maximally correlated classical state $\sigma_{A'B}=\frac{1}{d}\sum_{i=0}^{d-1}\dyad{i}_{A'}\otimes\dyad{i}_B.$ Let the generalized catalysis unitary $U$ acting on $AA'B$ be given as $U=\sum_{ijk} \dyad{j}{i \oplus 2k}_A \otimes \dyad{i \oplus j \oplus k}{i}_{A'} \otimes \dyad{k}{i \oplus j}_B.$ Here, $\oplus$ denotes the addition modulo $d$. Another extremal example is, when $d=m^2$ for some integer $m$, a pair of $m$-dimensional maximally mixed states and a $m$-dimensional maximally entangled pure state, i.e. $\frac{1}{m}\mathds{1}_{A_1}\otimes\dyad{\Psi}_{A_2B_2}\otimes \frac{1}{m}\mathds{1}_{B_1}.$ The generalized catalysis unitary consists of multiple steps. First, assign an arbitrary bipartite structure to the input system $A$ and swap it with system $A_2B_2$. Next, mask the system $A_2$ by using $B_1$ as a randomness source and similarly mask the system $B_2$ by using $A_1$ as a randomness source. Examples of masking unitaries are given Ref. \cite{lie2019unconditionally}.

Proposition \ref{prop:initial} suggests that entropy absorption of quantum catalyst shows the dual behavior. Even if the entropy is locally decreased by catalysis, when a reference system with which the input system is correlated is introduced, the entropy of the reference-input joint system can increase. We will call this increase of entropy the global increase of entropy. Therefore an intermediate should not only have enough free randomness but also enough room to absorb external randomness. This observation can be generalized to the following Theorem.

\begin{theoremEnd}{prop}
    Any quantum map that locally decreases entropy by $\Delta S$ should globally increases entropy by at least $\Delta S$.
\end{theoremEnd}

\begin{proofEnd}
    Let $\Phi$ be the quantum map in question and let $A$ be the system $\Phi$ acts on. Let $\gamma$ be the quantum state that achieves the entropy decrease of $\Delta S$, i.e. $S(\gamma)-S(\Phi(\gamma))=\Delta S$. Consider a purification $\ket{G}_{AB}$ of $\gamma$, i.e. $\Tr_B\dyad{G}_{AB}=\gamma_A$. Next, we let $\zeta_{AB}:=(\Phi_A\otimes\mathcal{I}_B)(\dyad{G}_{AB})$ and use the inequality $S(B)_\zeta - S(A)_\zeta \leq S(AB)_\zeta$ from the Araki-Lieb inequality of the von Neumann entropy \cite{araki1970entropy}. Note that $S(A)_\zeta=S(\Phi(\gamma))$ and $S(B)_\zeta=S(\gamma)$. Therefore $S(B)_\zeta - S(A)_\zeta$ equals to the local decrease of entropy by $\Phi$. Similarly, $S(AB)_\zeta$ can be interpreted as the gloval entropy increase of the pure input state $\ket{G}_{AB}$ as a pure state has zero von Neumann entropy. This proves the desired result.
\end{proofEnd}
This result shows that a maximally correlated intermediate $\sigma_{AB}$, i.e. $I(A:B)_\sigma=2S(B)_\sigma$, cannot be used for catalytical implementation of \textit{any} quantum map which causes entropy change.

\subsection{Secret-decoding map}
The no-hiding theorem \cite{braunstein2007quantum} can be restated as that the complementary channel of an erasure channel is an isometry. In other words, if quantum information completely disappears from a system, then it can be deterministically retrieved from its purifications. However, it is possible to circumvent the no-hiding theorem and hide the quantum information from local parties if we allow the initial state of the ancillary system to be mixed. Such a hiding process is equivalent to catalytic implementation of erasure channel.

Nevertheless, the following form of generalization of the no-hiding theorem applies to this situation, too \cite{lie2019unconditionally}. If the joint system $BC$ is in a pure state, then, when the whole quantum state of the system $A$ is encoded solely into the correlation of the joint system $AB$ (i.e. without altering the marginal state of $B$), it can be deterministically retrieved from the correlation of the joint system $AC$ too. That is, it is impossible to hide a quantum state into the correlation of only one pair of quantum systems, since there is always another system the correlation with which stores the hidden quantum state. It implies that quantum information cannot be localized in the correlation of a unique pair of quantum systems.

A further generalization of this result named \textit{the no-secret theorem}, which generalizes the complete erasure to arbitrary degrading of quantum information, was proved in Ref. \cite{lie2020uniform}. We introduce its proof here for completeness. Assume that a quantum map $\Phi$ on system $A$ is implemented through a generalized randomness-utilizing process, i.e. no information about the input state of $\Phi$ is leaked to the ancillary system other than the information that the map is implemented, with a unitary $M$ acting on $AB$ and a randomness source $\sigma$ in system $B$. $\sigma_B$ transforms into $\tau_B$ after the implementation, regardless of the input state.  Let $C$ be a purification system $\sigma_B$, i.e. $\sigma_{BC}$ is pure state such that $\Tr_C\sigma_{BC}=\sigma_B$. We input the part of a maximally entangled state $\Psi_{RA}$ into $\Phi$ and similarly consider a purification $\tau_{BC}$ of $\tau_B$. The marginal state on $RB$ is $\frac{1}{d}\mathds{1}_R\otimes \tau_B$, whose another purification is $\Psi_{RA}\otimes \tau_{BC}$. Since every purification of the same quantum state are unitarily similar on the purifying system, we acquire the existence of unitary operator $V$ acting on $AC$ such that $V_{AC}M_{AB}(\Psi_{RA}\otimes\sigma_{BC})M_{AB}^\dag V_{AC}^\dag = \Psi_{RA}\otimes \tau_{BC}$.

Considering the Choi-Jamiołkowski isomorphism, we can say that the information hidden between $A$ and $B$ by $M_{AB}$ can be restored  by the interaction between $A$ and $C$, i.e. $V_{AC}$. It shows that not only the whole quantum state, but also \textit{any} kind of quantum information encoded into the correlation of a pair of quantum systems must be able to be stored from an interaction between another pair of quantum systems. Note that the condition that no information should be leaked to a local system throughout the process is crucial. A localized information, of course, cannot be restored from another system unless it was copied beforehand.

\begin{theoremEnd}{theorem} [the no-secret theorem, \cite{lie2020uniform}] \label{thm:leakall}
    There is no way to confine quantum information in the correlation between a single pair of quantum systems.
\end{theoremEnd}
Theorem \ref{thm:leakall} can be understood as a quantum generalization of the fact that any information encrypted with a random variable $X$ as a key can be decrypted with any random variable $Y$ that is maximally correlated with $X$, i.e. $I(X:Y)=H(X)$. A remarkable point is that the encryption need not be perfect; Theorem $\ref{thm:leakall}$ applies to any encryption with arbitrary level of concealing.

Theorem \ref{thm:leakall}, however, merely implies the existence of the unitary operator that recovers the concealed quantum information. The characterization of catalysis unitary given by Theorem \ref{thm:catauni} shows what that unitary operator is. For a catalysis implemented with a catalysis unitary $U_{AB}$, the corresponding recovery map for the system $AC$ is the partial transpose $U_{AC}^{T_B}$, since $U_{AB}\kappa_A\otimes\sigma_{BC}U_{AB}^\dag=U_{AC}^{T_B}\kappa_A\otimes\sigma_{BC} U_{AC}^{T_B\dag}$ as $U_{AB}$ for any $\kappa_A$ commutes with $\sigma_B$ and $\sigma_{BC}=(\sqrt{\sigma_B}\otimes\mathds{1}_C)\dyad{\Gamma}_{BC}(\sqrt{\sigma_B}\otimes\mathds{1}_C)$ for an unnormalized maximally entangled state $\ket{\Gamma}_{BC}=\sum_i \ket{i}_B\ket{i}_C$.

\subsection{Advantage of explicit model}

A notable example of the advantage of adopting the explicit model of correlation being evident is the case where the intermediate $\sigma_{A_2B}$ is a classical-quantum state, i.e. $\sigma_{A_2B}=\sum_{i=1}^N p_i  \dyad{i}_{A_2}\otimes \dyad{\psi_i}_B$ with some probability distribution $\{p_i\}$. It is equivalent to the situation where a random pure state $\ket{\psi_i}$ is generated but the agent $A$ has the perfect knowledge of $B$ in the memory $A_2$. When the correlation with randomness source is treated implicitly, one may denote the state of the randomness source $B$ as a randomly chosen but pure state $\dyad{\psi_i}$ with no randomness at all, i.e. $S(\dyad{\psi_i})=0$. It will render the randomness source useless even when it is used quantum mechanically. However, if one adopts the explicit model of correlation, then, for the case $N=d$ and $p_i=\frac{1}{d}$ with $\ket{\psi_i}=\ket{i}$, we can see that the randomness source still has $2S(B)_\sigma - I(A_2:B)_\sigma=\log_2  d$ bits of free randomness. One can even erase $\log_2  d$ qubits of quantum information with this randomness source.

\subsection{Multi-party infinite catalysis}

We have seen that a catalyst has a limited power as a randomness source and once its randomness is depleted then it cannot be used for randomization. Does it mean that if the number of independent users of the same catalyst is finite, then the number of usages of the catalyst is limited? In the following, we introduce a counterexample to this hypothesis.

Suppose that there are two separated parties, $A$ and $B$, who wish to implement dephasing maps with respect to the computational basis $(i.e. \{\ket{i}\})$ on $d^2$-dimensional quantum systems using a catalyst $C$ in the state $\frac{1}{d}\mathds{1}_C$ using the method given in \cite{boes2018catalytic}. For her first turn, $A$ dephases a pure state that is unbiased to the computational basis, say, $\ket{+}=\frac{1}{d}\sum_{i=1}^{d^2} \ket{i}$. It results in the complete depletion of the randomness of the catalyst. After it, $A$ hands over the catalyst to $B$ and $B$ implements the same dephasing map upon the same, but independently prepared state $\ket{+}$. Again, the catalyst becomes exhausted for $B$. The total state of $ABC$, which we will call the joint-intermediate, has the following form at this stage.

\begin{equation}
  \tau_{ABC} =  \frac{1}{d^4}\sum_{ijkl}\dyad{i}{j}_A \otimes \dyad{k}{l}_B \otimes (U_k U_i U_j^\dag U_l^\dag)_C,
\end{equation}
Where $\{U_i\}$ is a set of orthonormal unitary operators, i.e. $\Tr U_i U_j^\dag=d\delta_{ij}$. However, when $B$ returns the catalyst back to $A$, from the perspective of $A$, the catalyst looks `refuelled'. It is because the marginal state on the system $AC$ decoupled, i.e.

\begin{equation}
    \tau_{AC}= \frac{1}{d^3} \mathds{1}_A \otimes \mathds{1}_C.
\end{equation}
The same logic applies to $A$, too. Therefore, if they repeat this process, they can implement dephasing maps indefinitely many times.

This initialization of randomness happens because the complete depletion of randomness by $B$, i.e. $I(B:C)_\tau=2\log_2  d$ leads to the complete decoupling of  $AC$, because of the information conservation law \cite{lie2020randomness}. To be concrete, the following conservation law holds for any 4-partite pure state $\xi_{WXYZ}$,
\begin{equation} \label{eqn:mono}
    2S(Y)_\xi=I(X:Y)_\xi+I(Y:WZ)_\xi.
\end{equation}
From the data-processing inequality \cite{lieb1973proof} $I(Y:WZ)_\xi \geq I(Y:Z)_\xi$, by ignoring $W$ we get the inequality $2S(Y)_\xi \geq I(X:Y)_\xi + I(Y:Z)_\xi$. We apply this inequality to the joint-intermediate $\tau_{ABC}$ with $C$ being the catalyst. If $B$ nearly-depletes the randomness, i.e. $I(B:C)_\tau \geq 2S(C)_\tau - \epsilon$, then the randomness for $A$ is nearly-perfectly restored, i.e. $I(A:C)_\tau \leq \epsilon$. Note that obviously multiple users become more and more correlated as the usage of catalyst by them repeats.

The possibility of infinite catalysis with a finite number of users is a stark difference between quantum and classical catalyst. If $C$ is a classical catalyst, then the upper bound $I(B:C)_\tau \leq S(C)$ forbids the monogamous argument that upper bounds the mutual information of $I(A:C)_\tau$. Indeed, as any catalysis with a classical catalyst completely preserves the each eigenstate of the catalyst, the usage of the catalyst by other agents does not alter the intermediate of an agent at all. An agent cannot use the same catalyst twice.

\subsection{Catalytic implementation of\\ state transition vs. quantum map}
Previous studies on catalytic quantum randomness mainly focused on the transition between two specific quantum states with a correspondingly prepared catalyst. On the contrary, our main interest in this work is the implementation of quantum maps, not state transition. The former approach is highly effective for characterizing fundamental properties or the conditions for state transition. For example, it was newly discovered that the von Neumann entropy emerges among the family of R\'{e}nyi entropies as the only deciding factor if the catalytic transition between two specific states is possible, as the catalytic entropy conjecture, which was conjectured in Ref. \cite{boes2019neumann}, was recently proved by Wilming \cite{wilming2020entropy} using the technique introduced by Shiraishi and Sagawa \cite{shiraishi2020quantum}.

The aforementioned technique is preparing a fine-tuned catalyst that is highly dependent on the initial and final states of the state transition in question. This setting, although it saturates the ultimate limit, is rather contrived from the operational perspective. It is because, since one needs different catalyst for each input and output state pair, one requires an enormous size of arsenal of catalysts for variable input and output state pair, which can easily be infinite. One should not need a different type of stove for cooking each dish; a tool must have a certain degree of versatility. If one assumes that a catalyst is built whenever it is required, one encounters a circular argument. How could one make a catalyst if randomness is not free? Therefore, it is more natural to treat a catalyst as a tool that takes resources to build and that one needs to return in its original form after every use.

In this setting, one starts with a given catalyst and the target map and input states can be decided afterwards with the capability of the catalyst in mind. This is the motivation of studying the catalytic implementation of quantum maps, instead of state transitions between specific states. In that case, it is logical to assess the power of a given catalyst, which was done in this work by finding the catalytic entropy of a given catalyst.

The most typical example to which this resource theory can be applied is the dephasing map. Dephasing map was shown to be catalytically implementable \cite{boes2018catalytic} and can be used for implementing state transition between two arbitrary quantum state, e.g. $\rho \to \rho'$ with majorization relation $\rho \succ \rho'$ by (quantum) Schur-Horn lemma \cite{scharlau2018quantum, horn1954doubly, schur1923uber, wilming2020entropy}. This type of usage of catalyst is not input-dependent, therefore subject to the resource theory of this work. For example, even if one tries to dephase almost-dephased input state $0.001 \dyad{+}+0.999\frac{1}{d}\mathds{1}$ (here, $\ket{+}=\frac{1}{\sqrt{d}}\sum_{i=1}^d\ket{i}$) to transform it into the maximally mixed state $\frac{1}{d}\mathds{1}$ with the catalyst $\frac{1}{d}\mathds{1}$, one cannot implement more than two times of the state transition of this type by naively implementing the tensor product of two dephasing maps, as its maximal entropy production exceeds the catalytic entropy of the catalyst, even if that entropy production does not actually take place.  In this sense, the resource theory of randomness for quantum maps is relevant to that for state transition.

\section{Conclusion and open problems}
We have seen that the maximally extractable randomness from an arbitrary mixed quantum state depends on the degeneracy of the state and can be quantified by the measure we defined in this work, the catalytic entropy. We highlighted an often overlooked fact that forming correlation with a catalyst depletes the useful randomness within it, by explicitly treating the correlation as a bipartite quantum state. We also gave an operational meaning associated with the partial transpose of bipartite unitary operators and showed that it works as the recovery operator of a catalysis unitary whose existence is guaranteed by the no-secret theorem.

This work opens up a broad field of research. We obtained a characterization of catalysis unitary for initially decoupled catalyst, however, the characterization for initially correlated catalyst is still an open problem. A second open problem is to find the `catalytic entropy of formation' of quantum maps, i.e. for a quantum map $\mathcal{N}$, find $S_{\alpha}^F(\mathcal{N}):=\min_{\sigma}{S_\alpha^\diamond(\sigma)}$ where the minimization is over the catalysts that can be used for catalytic implementation of $\mathcal{N}$. As the maximal entropy production of channel can be understood as the counterpart of distillable entanglement of entanglement theory, it is intuitive that $S_\alpha^G(\mathcal{N})\leq S_{\alpha}^F(\mathcal{N})$ holds. A natural conjecture is $S_\alpha^G(\mathcal{N})=S_{\alpha}^F(\mathcal{N})$, but it would be interesting if it turns out that is not the case. 

\begin{acknowledgements}
This work was supported by National Research Foundation of Korea grants funded by the Korea government (Grants No. 2019M3E4A1080074, No. 2020R1A2C1008609 and No. 2020K2A9A1A06102946) via the Institute of Applied Physics at Seoul National University and by the Ministry of Science and ICT, Korea, under the ITRC (Information Technology Research Center) support program (IITP-2020-0-01606) supervised by the IITP (Institute of Information \& Communications Technology Planning \& Evaluation).
\end{acknowledgements}

\appendix
\section{Proofs of results} \label{appendix}
\printProofs

\bibliography{main}

\begin{thebibliography}{31}%
\makeatletter
\providecommand \@ifxundefined [1]{%
 \@ifx{#1\undefined}
}%
\providecommand \@ifnum [1]{%
 \ifnum #1\expandafter \@firstoftwo
 \else \expandafter \@secondoftwo
 \fi
}%
\providecommand \@ifx [1]{%
 \ifx #1\expandafter \@firstoftwo
 \else \expandafter \@secondoftwo
 \fi
}%
\providecommand \natexlab [1]{#1}%
\providecommand \enquote  [1]{``#1''}%
\providecommand \bibnamefont  [1]{#1}%
\providecommand \bibfnamefont [1]{#1}%
\providecommand \citenamefont [1]{#1}%
\providecommand \href@noop [0]{\@secondoftwo}%
\providecommand \href [0]{\begingroup \@sanitize@url \@href}%
\providecommand \@href[1]{\@@startlink{#1}\@@href}%
\providecommand \@@href[1]{\endgroup#1\@@endlink}%
\providecommand \@sanitize@url [0]{\catcode `\\12\catcode `\$12\catcode
  `\&12\catcode `\#12\catcode `\^12\catcode `\_12\catcode `\%12\relax}%
\providecommand \@@startlink[1]{}%
\providecommand \@@endlink[0]{}%
\providecommand \url  [0]{\begingroup\@sanitize@url \@url }%
\providecommand \@url [1]{\endgroup\@href {#1}{\urlprefix }}%
\providecommand \urlprefix  [0]{URL }%
\providecommand \Eprint [0]{\href }%
\providecommand \doibase [0]{https://doi.org/}%
\providecommand \selectlanguage [0]{\@gobble}%
\providecommand \bibinfo  [0]{\@secondoftwo}%
\providecommand \bibfield  [0]{\@secondoftwo}%
\providecommand \translation [1]{[#1]}%
\providecommand \BibitemOpen [0]{}%
\providecommand \bibitemStop [0]{}%
\providecommand \bibitemNoStop [0]{.\EOS\space}%
\providecommand \EOS [0]{\spacefactor3000\relax}%
\providecommand \BibitemShut  [1]{\csname bibitem#1\endcsname}%
\let\auto@bib@innerbib\@empty
\bibitem [{\citenamefont {Boes}\ \emph {et~al.}(2018)\citenamefont {Boes},
  \citenamefont {Wilming}, \citenamefont {Gallego},\ and\ \citenamefont
  {Eisert}}]{boes2018catalytic}%
  \BibitemOpen
  \bibfield  {author} {\bibinfo {author} {\bibfnamefont {P.}~\bibnamefont
  {Boes}}, \bibinfo {author} {\bibfnamefont {H.}~\bibnamefont {Wilming}},
  \bibinfo {author} {\bibfnamefont {R.}~\bibnamefont {Gallego}},\ and\ \bibinfo
  {author} {\bibfnamefont {J.}~\bibnamefont {Eisert}},\ }\bibfield  {title}
  {\bibinfo {title} {Catalytic quantum randomness},\ }\href@noop {} {\bibfield
  {journal} {\bibinfo  {journal} {Physical Review X}\ }\textbf {\bibinfo
  {volume} {8}},\ \bibinfo {pages} {041016} (\bibinfo {year}
  {2018})}\BibitemShut {NoStop}%
\bibitem [{\citenamefont {Boes}\ \emph {et~al.}(2019)\citenamefont {Boes},
  \citenamefont {Eisert}, \citenamefont {Gallego}, \citenamefont {M{\"u}ller},\
  and\ \citenamefont {Wilming}}]{boes2019neumann}%
  \BibitemOpen
  \bibfield  {author} {\bibinfo {author} {\bibfnamefont {P.}~\bibnamefont
  {Boes}}, \bibinfo {author} {\bibfnamefont {J.}~\bibnamefont {Eisert}},
  \bibinfo {author} {\bibfnamefont {R.}~\bibnamefont {Gallego}}, \bibinfo
  {author} {\bibfnamefont {M.~P.}\ \bibnamefont {M{\"u}ller}},\ and\ \bibinfo
  {author} {\bibfnamefont {H.}~\bibnamefont {Wilming}},\ }\bibfield  {title}
  {\bibinfo {title} {Von neumann entropy from unitarity},\ }\href@noop {}
  {\bibfield  {journal} {\bibinfo  {journal} {Physical Review Letters}\
  }\textbf {\bibinfo {volume} {122}},\ \bibinfo {pages} {210402} (\bibinfo
  {year} {2019})}\BibitemShut {NoStop}%
\bibitem [{\citenamefont {Lie}\ \emph {et~al.}(2019)\citenamefont {Lie},
  \citenamefont {Kwon}, \citenamefont {Kim},\ and\ \citenamefont
  {Jeong}}]{lie2019unconditionally}%
  \BibitemOpen
  \bibfield  {author} {\bibinfo {author} {\bibfnamefont {S.~H.}\ \bibnamefont
  {Lie}}, \bibinfo {author} {\bibfnamefont {H.}~\bibnamefont {Kwon}}, \bibinfo
  {author} {\bibfnamefont {M.}~\bibnamefont {Kim}},\ and\ \bibinfo {author}
  {\bibfnamefont {H.}~\bibnamefont {Jeong}},\ }\bibfield  {title} {\bibinfo
  {title} {Unconditionally secure qubit commitment scheme using quantum
  maskers},\ }\href@noop {} {\bibfield  {journal} {\bibinfo  {journal} {arXiv
  preprint arXiv:1903.12304}\ } (\bibinfo {year} {2019})}\BibitemShut {NoStop}%
\bibitem [{\citenamefont {Lie}\ \emph {et~al.}(2020)\citenamefont {Lie},
  \citenamefont {Choi},\ and\ \citenamefont {Jeong}}]{lie2020minent}%
  \BibitemOpen
  \bibfield  {author} {\bibinfo {author} {\bibfnamefont {S.~H.}\ \bibnamefont
  {Lie}}, \bibinfo {author} {\bibfnamefont {S.}~\bibnamefont {Choi}},\ and\
  \bibinfo {author} {\bibfnamefont {H.}~\bibnamefont {Jeong}},\ }\bibfield
  {title} {\bibinfo {title} {The min-entropy as a resource for one-shot private
  state transfer, quantum masking and state transition},\ }\href@noop {}
  {\bibfield  {journal} {\bibinfo  {journal} {arXiv preprint arXiv:2010.14796}\
  } (\bibinfo {year} {2020})}\BibitemShut {NoStop}%
\bibitem [{\citenamefont {Lie}\ and\ \citenamefont
  {Jeong}(2020{\natexlab{a}})}]{lie2020randomness}%
  \BibitemOpen
  \bibfield  {author} {\bibinfo {author} {\bibfnamefont {S.~H.}\ \bibnamefont
  {Lie}}\ and\ \bibinfo {author} {\bibfnamefont {H.}~\bibnamefont {Jeong}},\
  }\bibfield  {title} {\bibinfo {title} {Randomness cost of masking quantum
  information and the information conservation law},\ }\href@noop {} {\bibfield
   {journal} {\bibinfo  {journal} {Physical Review A}\ }\textbf {\bibinfo
  {volume} {101}},\ \bibinfo {pages} {052322} (\bibinfo {year}
  {2020}{\natexlab{a}})}\BibitemShut {NoStop}%
\bibitem [{\citenamefont {M{\"u}ller}(2018)}]{muller2018correlating}%
  \BibitemOpen
  \bibfield  {author} {\bibinfo {author} {\bibfnamefont {M.~P.}\ \bibnamefont
  {M{\"u}ller}},\ }\bibfield  {title} {\bibinfo {title} {Correlating thermal
  machines and the second law at the nanoscale},\ }\href@noop {} {\bibfield
  {journal} {\bibinfo  {journal} {Physical Review X}\ }\textbf {\bibinfo
  {volume} {8}},\ \bibinfo {pages} {041051} (\bibinfo {year}
  {2018})}\BibitemShut {NoStop}%
\bibitem [{\citenamefont {R{\'e}nyi}\ \emph {et~al.}(1961)\citenamefont
  {R{\'e}nyi} \emph {et~al.}}]{renyi1961measures}%
  \BibitemOpen
  \bibfield  {author} {\bibinfo {author} {\bibfnamefont {A.}~\bibnamefont
  {R{\'e}nyi}} \emph {et~al.},\ }\bibfield  {title} {\bibinfo {title} {On
  measures of entropy and information},\ }in\ \href@noop {} {\emph {\bibinfo
  {booktitle} {Proceedings of the Fourth Berkeley Symposium on Mathematical
  Statistics and Probability, Volume 1: Contributions to the Theory of
  Statistics}}}\ (\bibinfo {organization} {The Regents of the University of
  California},\ \bibinfo {year} {1961})\BibitemShut {NoStop}%
\bibitem [{\citenamefont {Horodecki}\ \emph {et~al.}(2003)\citenamefont
  {Horodecki}, \citenamefont {Horodecki},\ and\ \citenamefont
  {Oppenheim}}]{horodecki2003reversible}%
  \BibitemOpen
  \bibfield  {author} {\bibinfo {author} {\bibfnamefont {M.}~\bibnamefont
  {Horodecki}}, \bibinfo {author} {\bibfnamefont {P.}~\bibnamefont
  {Horodecki}},\ and\ \bibinfo {author} {\bibfnamefont {J.}~\bibnamefont
  {Oppenheim}},\ }\bibfield  {title} {\bibinfo {title} {Reversible
  transformations from pure to mixed states and the unique measure of
  information},\ }\href@noop {} {\bibfield  {journal} {\bibinfo  {journal}
  {Physical Review A}\ }\textbf {\bibinfo {volume} {67}},\ \bibinfo {pages}
  {062104} (\bibinfo {year} {2003})}\BibitemShut {NoStop}%
\bibitem [{\citenamefont {Scharlau}\ and\ \citenamefont
  {Mueller}(2018)}]{scharlau2018quantum}%
  \BibitemOpen
  \bibfield  {author} {\bibinfo {author} {\bibfnamefont {J.}~\bibnamefont
  {Scharlau}}\ and\ \bibinfo {author} {\bibfnamefont {M.~P.}\ \bibnamefont
  {Mueller}},\ }\bibfield  {title} {\bibinfo {title} {Quantum horn's lemma,
  finite heat baths, and the third law of thermodynamics},\ }\href@noop {}
  {\bibfield  {journal} {\bibinfo  {journal} {Quantum}\ }\textbf {\bibinfo
  {volume} {2}},\ \bibinfo {pages} {54} (\bibinfo {year} {2018})}\BibitemShut
  {NoStop}%
\bibitem [{\citenamefont {Gour}\ \emph {et~al.}(2015)\citenamefont {Gour},
  \citenamefont {M{\"u}ller}, \citenamefont {Narasimhachar}, \citenamefont
  {Spekkens},\ and\ \citenamefont {Halpern}}]{gour2015resource}%
  \BibitemOpen
  \bibfield  {author} {\bibinfo {author} {\bibfnamefont {G.}~\bibnamefont
  {Gour}}, \bibinfo {author} {\bibfnamefont {M.~P.}\ \bibnamefont
  {M{\"u}ller}}, \bibinfo {author} {\bibfnamefont {V.}~\bibnamefont
  {Narasimhachar}}, \bibinfo {author} {\bibfnamefont {R.~W.}\ \bibnamefont
  {Spekkens}},\ and\ \bibinfo {author} {\bibfnamefont {N.~Y.}\ \bibnamefont
  {Halpern}},\ }\bibfield  {title} {\bibinfo {title} {The resource theory of
  informational nonequilibrium in thermodynamics},\ }\href@noop {} {\bibfield
  {journal} {\bibinfo  {journal} {Physics Reports}\ }\textbf {\bibinfo {volume}
  {583}},\ \bibinfo {pages} {1} (\bibinfo {year} {2015})}\BibitemShut {NoStop}%
\bibitem [{\citenamefont {Lie}\ and\ \citenamefont
  {Jeong}(2020{\natexlab{b}})}]{lie2020uniform}%
  \BibitemOpen
  \bibfield  {author} {\bibinfo {author} {\bibfnamefont {S.~H.}\ \bibnamefont
  {Lie}}\ and\ \bibinfo {author} {\bibfnamefont {H.}~\bibnamefont {Jeong}},\
  }\bibfield  {title} {\bibinfo {title} {Only uniform randomness can yield
  quantum advantages},\ }\href@noop {} {\bibfield  {journal} {\bibinfo
  {journal} {arXiv preprint arXiv:2010.14795}\ } (\bibinfo {year}
  {2020}{\natexlab{b}})}\BibitemShut {NoStop}%
\bibitem [{\citenamefont {Buscemi}(2009)}]{buscemi2009private}%
  \BibitemOpen
  \bibfield  {author} {\bibinfo {author} {\bibfnamefont {F.}~\bibnamefont
  {Buscemi}},\ }\bibfield  {title} {\bibinfo {title} {Private quantum
  decoupling and secure disposal of information},\ }\href@noop {} {\bibfield
  {journal} {\bibinfo  {journal} {New Journal of Physics}\ }\textbf {\bibinfo
  {volume} {11}},\ \bibinfo {pages} {123002} (\bibinfo {year}
  {2009})}\BibitemShut {NoStop}%
\bibitem [{\citenamefont {Gottesman}(2000)}]{gottesman2000theory}%
  \BibitemOpen
  \bibfield  {author} {\bibinfo {author} {\bibfnamefont {D.}~\bibnamefont
  {Gottesman}},\ }\bibfield  {title} {\bibinfo {title} {Theory of quantum
  secret sharing},\ }\href@noop {} {\bibfield  {journal} {\bibinfo  {journal}
  {Physical Review A}\ }\textbf {\bibinfo {volume} {61}},\ \bibinfo {pages}
  {042311} (\bibinfo {year} {2000})}\BibitemShut {NoStop}%
\bibitem [{\citenamefont {Cleve}\ \emph {et~al.}(1999)\citenamefont {Cleve},
  \citenamefont {Gottesman},\ and\ \citenamefont {Lo}}]{cleve1999share}%
  \BibitemOpen
  \bibfield  {author} {\bibinfo {author} {\bibfnamefont {R.}~\bibnamefont
  {Cleve}}, \bibinfo {author} {\bibfnamefont {D.}~\bibnamefont {Gottesman}},\
  and\ \bibinfo {author} {\bibfnamefont {H.-K.}\ \bibnamefont {Lo}},\
  }\bibfield  {title} {\bibinfo {title} {How to share a quantum secret},\
  }\href@noop {} {\bibfield  {journal} {\bibinfo  {journal} {Physical Review
  Letters}\ }\textbf {\bibinfo {volume} {83}},\ \bibinfo {pages} {648}
  (\bibinfo {year} {1999})}\BibitemShut {NoStop}%
\bibitem [{\citenamefont {Imai}\ \emph {et~al.}(2005)\citenamefont {Imai},
  \citenamefont {M{\"u}ller-Quade}, \citenamefont {Nascimento}, \citenamefont
  {Tuyls},\ and\ \citenamefont {Winter}}]{imai2005information}%
  \BibitemOpen
  \bibfield  {author} {\bibinfo {author} {\bibfnamefont {H.}~\bibnamefont
  {Imai}}, \bibinfo {author} {\bibfnamefont {J.}~\bibnamefont
  {M{\"u}ller-Quade}}, \bibinfo {author} {\bibfnamefont {A.~C.}\ \bibnamefont
  {Nascimento}}, \bibinfo {author} {\bibfnamefont {P.}~\bibnamefont {Tuyls}},\
  and\ \bibinfo {author} {\bibfnamefont {A.}~\bibnamefont {Winter}},\
  }\bibfield  {title} {\bibinfo {title} {An information theoretical model for
  quantum secret sharing},\ }\href@noop {} {\bibfield  {journal} {\bibinfo
  {journal} {Quantum Information \& Computation}\ }\textbf {\bibinfo {volume}
  {5}},\ \bibinfo {pages} {69} (\bibinfo {year} {2005})}\BibitemShut {NoStop}%
\bibitem [{\citenamefont {Shor}(2010)}]{shor2010structure}%
  \BibitemOpen
  \bibfield  {author} {\bibinfo {author} {\bibfnamefont {P.}~\bibnamefont
  {Shor}},\ }\href@noop {} {\bibinfo {title} {Structure of unital maps and the
  asymptotic quantum birkhoff conjecture}} (\bibinfo {year} {2010})\BibitemShut
  {NoStop}%
\bibitem [{\citenamefont {Haagerup}\ and\ \citenamefont
  {Musat}(2011)}]{haagerup2011factorization}%
  \BibitemOpen
  \bibfield  {author} {\bibinfo {author} {\bibfnamefont {U.}~\bibnamefont
  {Haagerup}}\ and\ \bibinfo {author} {\bibfnamefont {M.}~\bibnamefont
  {Musat}},\ }\bibfield  {title} {\bibinfo {title} {Factorization and dilation
  problems for completely positive maps on von neumann algebras},\ }\href@noop
  {} {\bibfield  {journal} {\bibinfo  {journal} {Communications in Mathematical
  Physics}\ }\textbf {\bibinfo {volume} {303}},\ \bibinfo {pages} {555}
  (\bibinfo {year} {2011})}\BibitemShut {NoStop}%
\bibitem [{\citenamefont {Deschamps}\ \emph {et~al.}(2016)\citenamefont
  {Deschamps}, \citenamefont {Nechita},\ and\ \citenamefont
  {Pellegrini}}]{deschamps2016some}%
  \BibitemOpen
  \bibfield  {author} {\bibinfo {author} {\bibfnamefont {J.}~\bibnamefont
  {Deschamps}}, \bibinfo {author} {\bibfnamefont {I.}~\bibnamefont {Nechita}},\
  and\ \bibinfo {author} {\bibfnamefont {C.}~\bibnamefont {Pellegrini}},\
  }\bibfield  {title} {\bibinfo {title} {On some classes of bipartite unitary
  operators},\ }\href@noop {} {\bibfield  {journal} {\bibinfo  {journal}
  {Journal of Physics A: Mathematical and Theoretical}\ }\textbf {\bibinfo
  {volume} {49}},\ \bibinfo {pages} {335301} (\bibinfo {year}
  {2016})}\BibitemShut {NoStop}%
\bibitem [{\citenamefont {Benoist}\ and\ \citenamefont
  {Nechita}(2017)}]{benoist2017bipartite}%
  \BibitemOpen
  \bibfield  {author} {\bibinfo {author} {\bibfnamefont {T.}~\bibnamefont
  {Benoist}}\ and\ \bibinfo {author} {\bibfnamefont {I.}~\bibnamefont
  {Nechita}},\ }\bibfield  {title} {\bibinfo {title} {On bipartite unitary
  matrices generating subalgebra-preserving quantum operations},\ }\href@noop
  {} {\bibfield  {journal} {\bibinfo  {journal} {Linear Algebra and its
  Applications}\ }\textbf {\bibinfo {volume} {521}},\ \bibinfo {pages} {70}
  (\bibinfo {year} {2017})}\BibitemShut {NoStop}%
\bibitem [{\citenamefont {Araki}\ and\ \citenamefont
  {Lieb}(1970)}]{araki1970entropy}%
  \BibitemOpen
  \bibfield  {author} {\bibinfo {author} {\bibfnamefont {H.}~\bibnamefont
  {Araki}}\ and\ \bibinfo {author} {\bibfnamefont {E.~H.}\ \bibnamefont
  {Lieb}},\ }\bibfield  {title} {\bibinfo {title} {Entropy inequalities},\
  }\href@noop {} {\bibfield  {journal} {\bibinfo  {journal} {Communications in
  Mathematical Physics}\ }\textbf {\bibinfo {volume} {18}},\ \bibinfo {pages}
  {160} (\bibinfo {year} {1970})}\BibitemShut {NoStop}%
\bibitem [{\citenamefont {Groisman}\ \emph {et~al.}(2005)\citenamefont
  {Groisman}, \citenamefont {Popescu},\ and\ \citenamefont
  {Winter}}]{groisman2005quantum}%
  \BibitemOpen
  \bibfield  {author} {\bibinfo {author} {\bibfnamefont {B.}~\bibnamefont
  {Groisman}}, \bibinfo {author} {\bibfnamefont {S.}~\bibnamefont {Popescu}},\
  and\ \bibinfo {author} {\bibfnamefont {A.}~\bibnamefont {Winter}},\
  }\bibfield  {title} {\bibinfo {title} {Quantum, classical, and total amount
  of correlations in a quantum state},\ }\href@noop {} {\bibfield  {journal}
  {\bibinfo  {journal} {Physical Review A}\ }\textbf {\bibinfo {volume} {72}},\
  \bibinfo {pages} {032317} (\bibinfo {year} {2005})}\BibitemShut {NoStop}%
\bibitem [{\citenamefont {Braunstein}\ and\ \citenamefont
  {Pati}(2007)}]{braunstein2007quantum}%
  \BibitemOpen
  \bibfield  {author} {\bibinfo {author} {\bibfnamefont {S.~L.}\ \bibnamefont
  {Braunstein}}\ and\ \bibinfo {author} {\bibfnamefont {A.~K.}\ \bibnamefont
  {Pati}},\ }\bibfield  {title} {\bibinfo {title} {Quantum information cannot
  be completely hidden in correlations: implications for the black-hole
  information paradox},\ }\href@noop {} {\bibfield  {journal} {\bibinfo
  {journal} {Physical review letters}\ }\textbf {\bibinfo {volume} {98}},\
  \bibinfo {pages} {080502} (\bibinfo {year} {2007})}\BibitemShut {NoStop}%
\bibitem [{\citenamefont {Lieb}\ and\ \citenamefont
  {Ruskai}(1973)}]{lieb1973proof}%
  \BibitemOpen
  \bibfield  {author} {\bibinfo {author} {\bibfnamefont {E.~H.}\ \bibnamefont
  {Lieb}}\ and\ \bibinfo {author} {\bibfnamefont {M.~B.}\ \bibnamefont
  {Ruskai}},\ }\bibfield  {title} {\bibinfo {title} {Proof of the strong
  subadditivity of quantum-mechanical entropy},\ }\href@noop {} {\bibfield
  {journal} {\bibinfo  {journal} {Journal of Mathematical Physics}\ }\textbf
  {\bibinfo {volume} {14}},\ \bibinfo {pages} {1938} (\bibinfo {year}
  {1973})}\BibitemShut {NoStop}%
\bibitem [{\citenamefont {Wilming}(2020)}]{wilming2020entropy}%
  \BibitemOpen
  \bibfield  {author} {\bibinfo {author} {\bibfnamefont {H.}~\bibnamefont
  {Wilming}},\ }\bibfield  {title} {\bibinfo {title} {Entropy and reversible
  catalysis},\ }\href@noop {} {\bibfield  {journal} {\bibinfo  {journal} {arXiv
  preprint arXiv:2012.05573}\ } (\bibinfo {year} {2020})}\BibitemShut {NoStop}%
\bibitem [{\citenamefont {Shiraishi}\ and\ \citenamefont
  {Sagawa}(2020)}]{shiraishi2020quantum}%
  \BibitemOpen
  \bibfield  {author} {\bibinfo {author} {\bibfnamefont {N.}~\bibnamefont
  {Shiraishi}}\ and\ \bibinfo {author} {\bibfnamefont {T.}~\bibnamefont
  {Sagawa}},\ }\bibfield  {title} {\bibinfo {title} {Quantum thermodynamics of
  correlated-catalytic state conversion at small-scale},\ }\href@noop {}
  {\bibfield  {journal} {\bibinfo  {journal} {arXiv preprint arXiv:2010.11036}\
  } (\bibinfo {year} {2020})}\BibitemShut {NoStop}%
\bibitem [{\citenamefont {Horn}(1954)}]{horn1954doubly}%
  \BibitemOpen
  \bibfield  {author} {\bibinfo {author} {\bibfnamefont {A.}~\bibnamefont
  {Horn}},\ }\bibfield  {title} {\bibinfo {title} {Doubly stochastic matrices
  and the diagonal of a rotation matrix},\ }\href@noop {} {\bibfield  {journal}
  {\bibinfo  {journal} {American Journal of Mathematics}\ }\textbf {\bibinfo
  {volume} {76}},\ \bibinfo {pages} {620} (\bibinfo {year} {1954})}\BibitemShut
  {NoStop}%
\bibitem [{\citenamefont {Schur}(1923)}]{schur1923uber}%
  \BibitemOpen
  \bibfield  {author} {\bibinfo {author} {\bibfnamefont {I.}~\bibnamefont
  {Schur}},\ }\bibfield  {title} {\bibinfo {title} {Uber eine klasse von
  mittelbildungen mit anwendungen auf die determinantentheorie},\ }\href@noop
  {} {\bibfield  {journal} {\bibinfo  {journal} {Sitzungsberichte der Berliner
  Mathematischen Gesellschaft}\ }\textbf {\bibinfo {volume} {22}},\ \bibinfo
  {pages} {51} (\bibinfo {year} {1923})}\BibitemShut {NoStop}%
\bibitem [{\citenamefont {Nielsen}\ and\ \citenamefont
  {Chuang}(2002)}]{nielsen2002quantum}%
  \BibitemOpen
  \bibfield  {author} {\bibinfo {author} {\bibfnamefont {M.~A.}\ \bibnamefont
  {Nielsen}}\ and\ \bibinfo {author} {\bibfnamefont {I.}~\bibnamefont
  {Chuang}},\ }\href@noop {} {\bibinfo {title} {Quantum computation and quantum
  information}} (\bibinfo {year} {2002})\BibitemShut {NoStop}%
\bibitem [{\citenamefont {Arias}\ \emph {et~al.}(2002)\citenamefont {Arias},
  \citenamefont {Gheondea},\ and\ \citenamefont {Gudder}}]{arias2002fixed}%
  \BibitemOpen
  \bibfield  {author} {\bibinfo {author} {\bibfnamefont {A.}~\bibnamefont
  {Arias}}, \bibinfo {author} {\bibfnamefont {A.}~\bibnamefont {Gheondea}},\
  and\ \bibinfo {author} {\bibfnamefont {S.}~\bibnamefont {Gudder}},\
  }\bibfield  {title} {\bibinfo {title} {Fixed points of quantum operations},\
  }\href@noop {} {\bibfield  {journal} {\bibinfo  {journal} {Journal of
  Mathematical Physics}\ }\textbf {\bibinfo {volume} {43}},\ \bibinfo {pages}
  {5872} (\bibinfo {year} {2002})}\BibitemShut {NoStop}%
\bibitem [{\citenamefont {van Dam}\ and\ \citenamefont
  {Hayden}(2002)}]{van2002renyi}%
  \BibitemOpen
  \bibfield  {author} {\bibinfo {author} {\bibfnamefont {W.}~\bibnamefont {van
  Dam}}\ and\ \bibinfo {author} {\bibfnamefont {P.}~\bibnamefont {Hayden}},\
  }\bibfield  {title} {\bibinfo {title} {Renyi-entropic bounds on quantum
  communication},\ }\href@noop {} {\bibfield  {journal} {\bibinfo  {journal}
  {arXiv preprint quant-ph/0204093}\ } (\bibinfo {year} {2002})}\BibitemShut
  {NoStop}%
\bibitem [{\citenamefont {Nielsen}(2000)}]{nielsen2000probability}%
  \BibitemOpen
  \bibfield  {author} {\bibinfo {author} {\bibfnamefont {M.~A.}\ \bibnamefont
  {Nielsen}},\ }\bibfield  {title} {\bibinfo {title} {Probability distributions
  consistent with a mixed state},\ }\href@noop {} {\bibfield  {journal}
  {\bibinfo  {journal} {Physical Review A}\ }\textbf {\bibinfo {volume} {62}},\
  \bibinfo {pages} {052308} (\bibinfo {year} {2000})}\BibitemShut {NoStop}%
\end{thebibliography}%

\end{document}